\documentclass[review]{elsarticle}

\usepackage{lineno,hyperref}
\modulolinenumbers[5]

\journal{arxiv.org}

\usepackage[export]{adjustbox} 

\usepackage{color,soul}
\usepackage{setspace}

\usepackage{subfig} 
\usepackage{amsmath}
\usepackage[english]{babel}
\usepackage{lmodern}
\usepackage[T1]{fontenc}
\usepackage{algorithmic}
\usepackage{fancyvrb}
\usepackage{booktabs}
\usepackage{graphicx}
\usepackage{graphics}
\usepackage{caption}
\usepackage{float}

 \usepackage{amssymb}
\usepackage{palatino}
\usepackage{amsmath}
\usepackage{longtable}    
\usepackage{graphicx}
\usepackage{xcolor}
\usepackage{url}
\usepackage{multicol}
\usepackage{multirow}
\usepackage{array}
\usepackage{lscape}
\usepackage{boxedminipage}
\usepackage{multicol}
\usepackage{multirow}
\usepackage{color,soul}

\begin{document}

\begin{frontmatter}


\title{A least-squares based nodal scheme for cell-centered Lagrangian hydrodynamics}

\author[]{Chuanjin Wang\corref{cor0}}
\ead{cwang35@ncsu.edu}
\author[]{Hong Luo\corref{cor1}}
\ead{hong\_luo@ncsu.edu}
\cortext[cor0]{Contribution from the first author was made when working at North Carolina State University.}

\address{
Department of Mechanical and Aerospace Engineering \\
North Carolina State University \\
Raleigh, NC 27695, USA}

\begin{abstract}
This paper presents a new and efficient nodal scheme for cell-centered compressible flows in Lagrangian formulation. 
A single pressure and the velocity at cell vertex are computed by the nodal solver in a least-squares sense, 
and both variables are used to evaluate the numerical flux across the cell interface. 
The resulting nodal velocity is also responsible for moving the mesh.
The accuracy and robustness of the proposed method is studied by several numerical examples in the finite volume discretization, 
and compared with two other nodal Riemann solvers. 
It is shown that its performance is comparable to the latter two. 
Although the current paper mainly focuses on the first-order finite volume (FV), 
its extension to higher order methods such as high-order FV, discontinuous Galerkin (DG) or reconstructed discontinous Galerkin (rDG),  is quite straightforward. 
And this method has the capability of easily extending to three dimensions. 
\end{abstract}


\end{frontmatter}


\section{Introduction}
Lagrangian method is widely used for flows undergoing large deformation, or for tracking multimaterial interface. 
The main concern in Lagrangian method is how to determine a unique velocity at cell vertex. 
The traditional way is to use staggered grid \cite{vonneumann1950sgh, wilkins1964sgh}, to place the variables at different grid locations.
This method has been studied extensively\cite{caramana1998sghConstruction, burton1994sgh,caramana1998sghElimination,caramana1998sghFormulations,campbell2001sghTensor,
lipnikov2010sghFramework,campbell2003sghCompatible} and become a robust and accurate method.

An alternative to the staggered-grid approach is the collocated-grid or cell-centered discretization, in which all the variables are defined at the cell-center. 
This method has gained wide attention and extensive exploration due to the consistency of locations for the variables and ease in implementation.
However, a difficulty stemmed in the cell-centered method is how to compute the vertex velocity, and ensure the consistency between the mesh motion and the numerical flux. 

Dukowicz et al.\cite{dukowicz1992vorticityError} and Addessio et al. \cite{CAVEAT} proposed an approximate one-directional Riemann solver in the face normal direction, and the cell-centered conserved quantities are then evolved using the resulting Riemann flux. 
One feature of this method is that the grid velocity is computed via a least-squares procedure concerning the face normal projection of the velocities, i.e., requiring the normal projection of the vertex velocity on each of the faces connecting to this vertex, to be equal to the Riemann face-normal velocity; this is to minimize the difference between these two.
This approach, unfortunately, will produce artificial mesh motion and inconsistency of the numerical flux with the mesh motion.
We should note that this least-squares procedure is a pure mathematical arithmetic treatment, \textit{after} solving the 1D face normal Riemann problems.

Cheng and Shu \cite{chengshu2007lag} developed a solver which also takes the advantage of the one-directional Riemann problem in the face normal direction.
To compute the vertex velocity, for example, for a vertex connecting four edges in a typical quadrilateral mesh, four velocity vectors are determined first, one for each of the four edges; then the final grid velocity of this vertex is a simple arithmetic average of the four vectors.
The velocity vector for each edge is determined as follows, separately for an edge \textit{normal} component and an edge \textit{tangential}  component.
At the vertex, we have the left state/velocity and right state/velocity, on two sides of the edge.
First the two velocity vectors are split into edge normal parts and tangential parts.
Then the edge \textit{tangential} component is set as the average of the two tangential parts; and the edge \textit{normal} component is solved by a one-directional Riemann solver in the edge normal direction, or by a Roe average, depending on the numerical flux scheme in use. 
Unfortunately, the numerical flux in this approach is still inconsistent with the mesh motion.

The evolution Galerkin type scheme \cite{prasad2001EG}, which is a multi-directional Riemann solver constructed at the cell vertex, can also be used for the Lagrangian methods. 
The evolution Galerkin solver in \cite{prasad2001EG} is designed based on the general theory of bicharacteristics, through exact integration of the linearized hyperbolic equations. 
This type of scheme has been investigated extensively \cite{lukacova2000EG,lukacova2002EG,lukacova2007EG,lukacova2009EG,
lukacova2011EG,wu2014EG,sunren2009euler,sunren2016lag}.
Sun and Ren \cite{sunren2009euler} devised a local evolution Galerkin solver for solving the compressible equations in Eulerian frame, and extended it to the Lagrangian formulation later by Sun et al \cite{sunren2016lag}. 
This local evolution Galerkin operator evolves the solution for an infinitely small time interval instead of the finite time-step used in the time marching. 
A good feature of this operator is that it decouples the temporal and spatial discretizations, while maintaining the multi-directional effect from the characteristic property of the compressible flow equations. 
For the Lagrangian frame in \cite{sunren2016lag}, the resulting vertex velocity and pressure from the local evolution Galerkin operator are used to update the grid coordinates and compute the numerical flux at cell interface consistently.
Although these evolution Galerkin schemes have impressive simulating capabilities, the extension to 3D is not so straightforward.

Another type of nodal solver, the nodal acoustic Riemann solver, is attracting more and more attention and research interest in the past decade.
Despr{\'e}s and Mazeran \cite{despres2005lag} proposed a multi-directional Riemann solver at the cell vertex for Lagrangian gas dynamics, in the context of finite volume discretization.
The momentum and total energy are conserved in this scheme, and an entropy inequality condition is ensured.
At a given node, a unique nodal Riemann velocity is defined at this node; one pressure per cell surrounding the node, is defined as the Riemann pressure, which will provide the forces and work, in the momentum and energy equations, respectively.
This constructs a consistent way to determine the vertex velocity and the numerical flux at the interface.
However, it was found that this scheme will lead to severe numerical instabilities, since the computed nodal velocity depends on the cell aspect ratio, even for the one-dimensional problem solved on a 2D mesh.

Maire et al. \cite{maire2007siam} investigated this issue and proposed an alternative Lagrangian scheme, which solves the aspect ratio problem, and inherits the consistent property in \cite{despres2005lag} between the nodal velocity and the numerical flux.
By construction, this solver recovers the classical Godunov Riemann solver in the one-dimensional case.
The main feature of this solver is the introduction of four pressures on each interface, two for each node on each side of the interface.
These pressures are connected to the unique nodal Riemann velocity by the Riemann jump relation.
The nodal Riemann velocity is solved by the assumption of a local equilibrium of the forces surrounding the node; this assumption is also an indication of the conservation of momentum and total energy.
Besides, the local entropy inequality is also satisfied in this scheme.

Burton et al. \cite{burton2013cchSolver} extended the seminal works of Despr{\'e}s and Mazeran \cite{despres2005lag} and  Maire et al. \cite{maire2007siam}, and proposed another robust multi-direcitonal Riemann nodal solver.
This node solver is capable of handling stress tensors, and is applied for materials with strength, e.g., elastic-plastic materials \cite{burton2013cchSolver}.
A good feature of this method is that the resulting Riemann stress tensors in the control volume corners are symmetric.
And a difference between this solver and those in \cite{despres2005lag} and \cite{maire2007siam} is that, the resulting Riemann force is always in the direction of the velocity difference (-- the difference between nodal Riemann velocity and the corner velocity of a cell), rather than in the face normal direction.

As is pointed out in \cite{sunren2016lag}, in the acoustic nodal Riemann solver\cite{maire2007siam}, the non-unique Riemann pressures at each interface leads to a nonequilibrium of numerical fluxes on two sides of the interface, and the sufficient conditions satisfied by each vertex for the local momentum and total energy conservation and the local entropy inequality are excessively strict.

In view of the advantageous properties of the both the evolution Galerkin type methods and the existing acoustic nodal Riemann solvers, 
the present paper proposes a new and efficient acoustic nodal solver, 
in which the unknowns are a single nodal Riemann pressure and the velocity components, and they are solved by a single least-squares problem derived from the Riemann jump equation.

The remainder of this paper is organized as follows.
The governing equations and the updated Lagrangian formulation  will follow in the next section.
Section \ref{sec:NodalSolver} introduces the new nodal Riemann solver.
Section \ref{sec:temporal} shows the temporal discretization.
A number of numerical examples are given in Section \ref{sec:example}.
Final conclusions are made in Section \ref{sec:conclusion}.

\section{Governing Equations}
The compressible Euler equations can be written in the vectorial form
\begin{equation}
\label{euler:eq:governing-equations}
  \frac{\partial\textbf{U}(\textbf{x},t)}{\partial t} 
  +\nabla \cdot \textbf{F}(\textbf{U})
= 0
\end{equation}
The conservative variable $\textbf{U}$ and the inviscid flux vector $\textbf{F}$ are defined as
\begin{equation}
\textbf{U}=
\begin{pmatrix} 
\rho \\ \rho \textbf{V} \\ \rho e 
\end{pmatrix}
\hspace{0.4in}
\textbf{F}=
\begin{pmatrix} 
\rho \textbf{V} \\ \rho \textbf{VV} + p\textbf{I} \\ (\rho e +p)\textbf{V} 
\end{pmatrix}
\end{equation}
where $\rho$, $p$ and $e$ denote the density, pressure and specific 
total energy of the fluid, respectively, and $\textbf{V}$ is the velocity vector of the flow field. 
The pressure can be computed from the equation of state
\begin{equation}
  p=(\gamma -1)\rho\left(e-\frac{1}{2}\lVert \textbf{V} \rVert^2 \right)  
\end{equation}
which is valid for perfect gas. 
$\gamma$ is the ratio of specific heats.

The unsteady compressible Euler equations for a moving control volume can be expressed in the unsplit ALE formulation
\begin{equation}
\begin{split}
\begin{aligned}
\frac{d}{dt}\int_{\Omega_e^t}\textbf{U} d\Omega
+
\int_{\Gamma_e^t}  \textbf{U}(\textbf{V}-\textbf{V}_g)\cdot \textbf{n} d\Gamma
+
\int_{\Gamma_e^t}  
\begin{pmatrix} 0 \\ p\textbf{I} \\ p\textbf{V} \end{pmatrix}
\cdot \textbf{n} d\Gamma
= 0
\end{aligned}
\end{split}
\end{equation}
where $\Omega_e^t$ is the moving control volume, $\Gamma_e^t$ its boundary, $\textbf{V}_g$ the arbitrary mesh velocity.
By assuming the fluid velocity equal to the grid velocity at cell boundaries, the above equations will reduce to the updated Lagrangian (or semi-Lagrangian) formulation 
\begin{equation}
\begin{split}
\begin{aligned}
\frac{d}{dt}\int_{\Omega_e^t}\textbf{U} d\Omega
+
\int_{\Gamma_e^t}  
\begin{pmatrix} 0 \\ p\textbf{I} \\ p\textbf{V} \end{pmatrix}
\cdot \textbf{n} d\Gamma
= 0
\end{aligned}
\end{split}
\end{equation}
In this paper, we use this formulation for the following Lagrangian computations.

\section{The Nodal Riemann Solver}
\label{sec:NodalSolver}

To move the mesh, we need to determine the mesh velocity at the vertices; to compute the numerical flux, we also need the Riemann pressure(s).
One main difference between the new solver in this paper (referred to as LS solver) and the other acoustic solvers (those by Maire et al.\cite{maire2007siam} and Burton et al. \cite{burton2013cchSolver}) is that, the new solver requires only one single Riemann pressure at the node, as illustrated in the figure below.
In this example, for a node surrounded by four cells, there will be 8, 4 and 1 Riemann pressure(s), respectively for these three solvers.

\begin{figure}[H]
  \centering
  \subfloat[Maire solver]{  
    \includegraphics[trim= 5cm 4cm 10cm 2cm,clip,height = 2in]
    {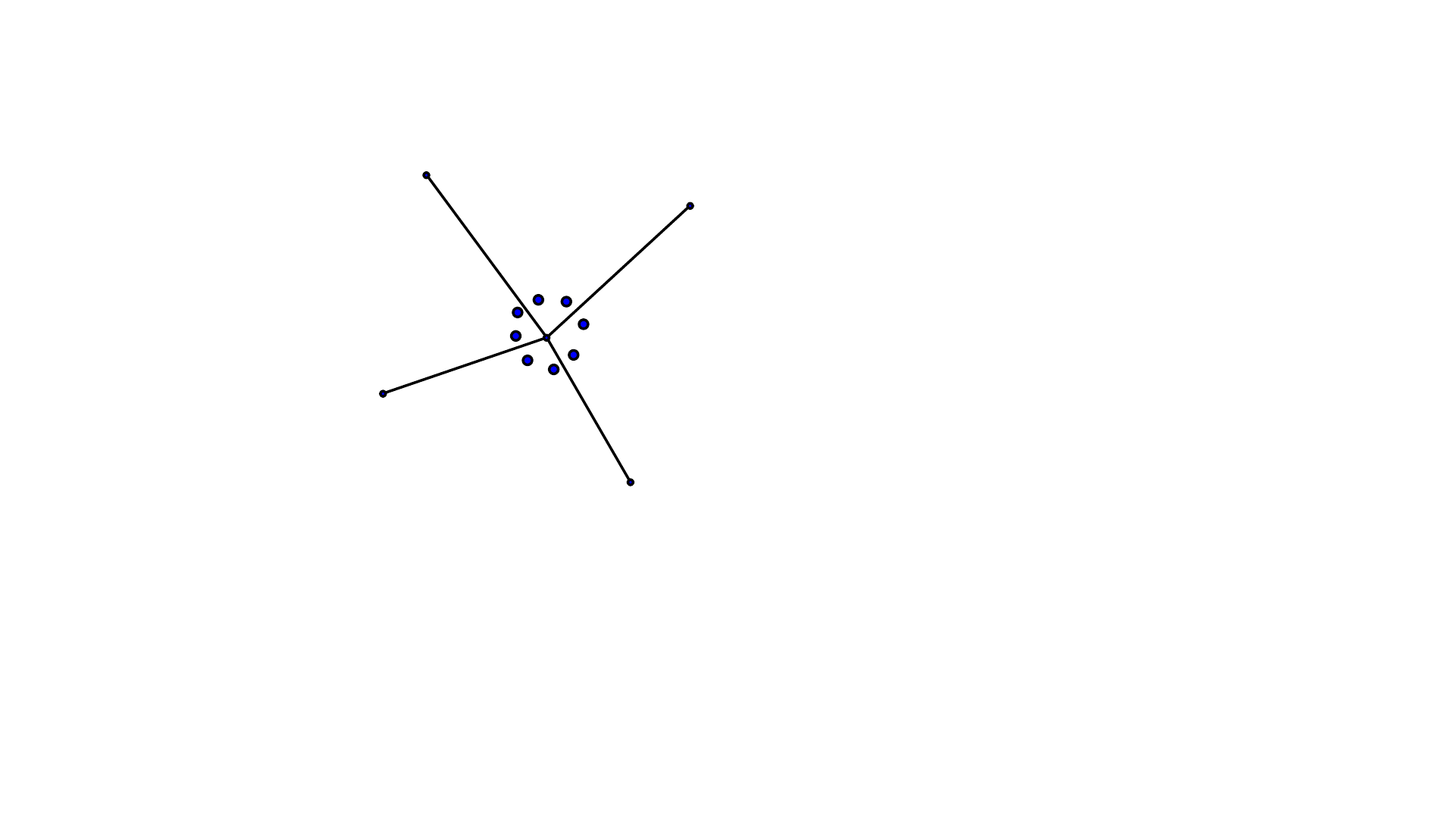} }
  \subfloat[Burton solver]{ 
    \includegraphics[trim= 5cm 4cm 10cm 2cm,clip,height = 2in]
    {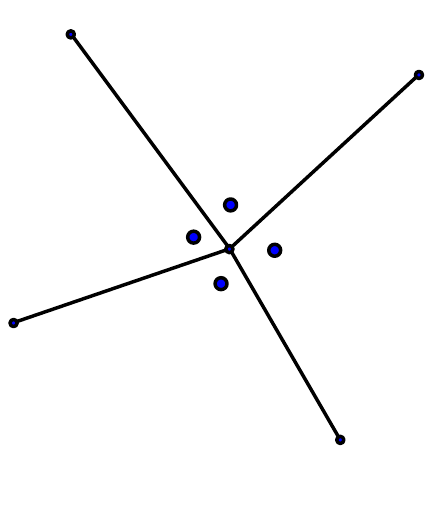} }
  \subfloat[current LS solver]{ 
    \includegraphics[trim= 5cm 4cm 10cm 2cm,clip,height = 2in]
    {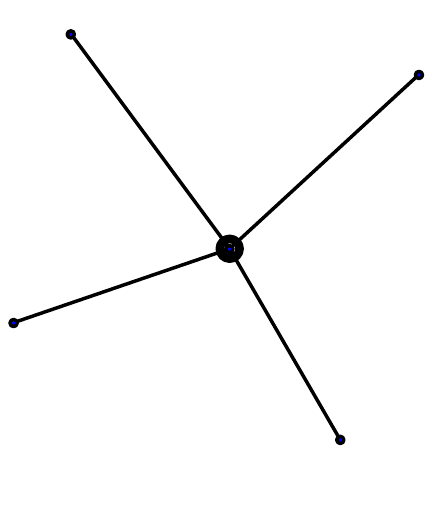} }    
  \caption{Number of Riemann pressures around a node.}
  \label{fig:solver:pressure}
\end{figure}
We first briefly recall the nodal solver by Maire et al.\cite{maire2007siam}, and then introduce the new LS solver proposed in this paper.

\subsection{The solver by Maire et al.\cite{maire2007siam}}
The main feature of this solver is the introduction of four pressures on each interface, two for each node on each side of the interface.
$\mathbf{F}_i^* = -p_i^*\mathbf{n}_i N_i$ is the Riemann pressure force acting on face segment $i$, where $p_i^*$ is the Riemann pressure and 
\begin{equation}
\label{eq:RiemannJump}
p_c - p_i^* = \mu_c
\left(\textbf{u}_p^* - \textbf{u}_c\right)
\cdot \textbf{n}_i
\end{equation}

\begin{figure}[H]
  \centering
  \includegraphics[trim= 5cm 2cm 7cm 1cm,clip,height =2.5in]    
    {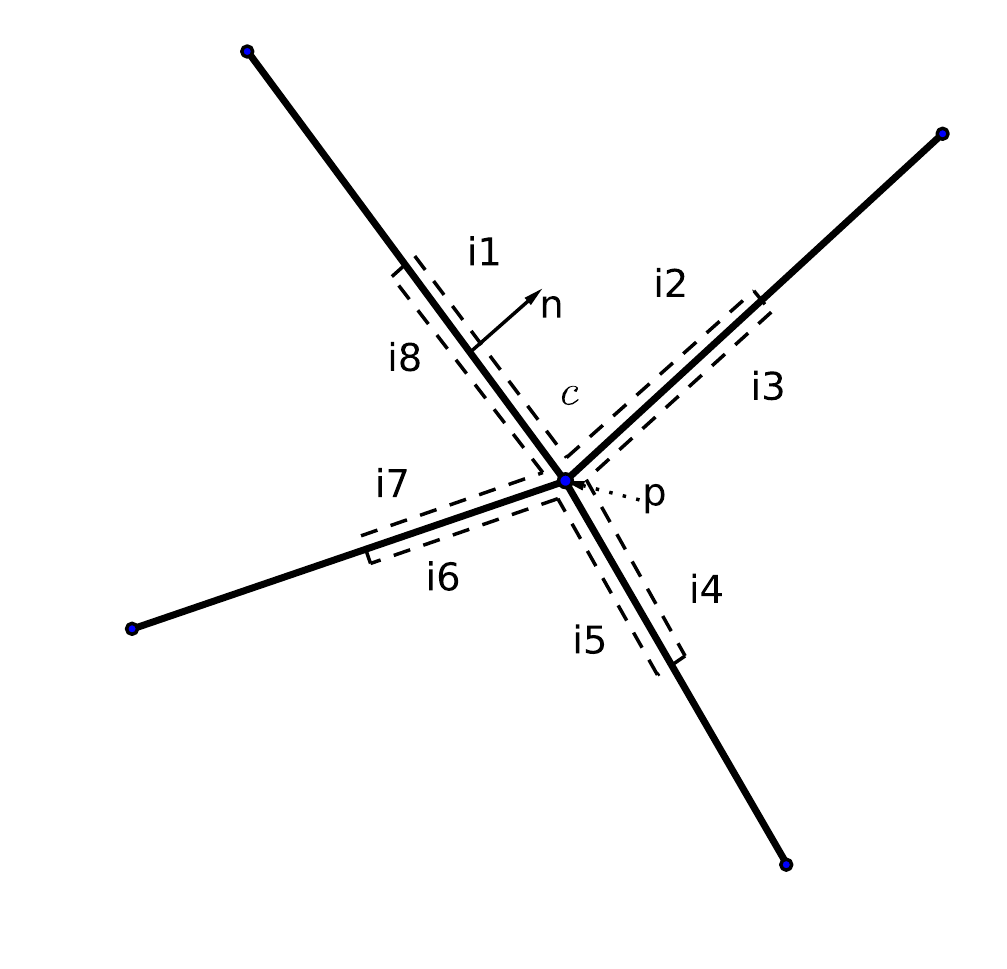}
  \caption{Notations of the nodal Riemann solver (Maire et al.\cite{maire2007siam}) at a node $p$.  
  The interface connected to $p$ is decomposed into a set of segments denoted by $i$, and $\textbf{n}_i$ is the outward face normal vector of the interface segment $i$.
  The subscript $c$ indicates which corner (cell) the segment $i$ belongs to.
  $\textbf{u}_p^*$ is the unique Riemann velocity defined at the node  and assumed to be constant over all the interface segments surrounding it.
$p_i^*$ is the Riemann pressure for each interface segment $i$.
  }
  \label{fig:nodeMaire}
\end{figure} 

In \cite{maire2007siam}, the acoustic impedance $\mu_c = \rho c$ is used , since the motivation was to recover the approximate acoustic solver for one-dimensional flows, where $c$ is the isentropic sound speed.

Then a local sufficient condition for the global conservation of momentum is 
\begin{equation}
\sum_{i\in p}\mathbf{F}_i^* = 0.
\end{equation}
This can be interpreted as the local equilibrium of node $p$ under pressure forces \cite{maire2007siam}. 
It will yield the equations for the Riemann velocity at the node
\begin{equation}
\begin{split}
\left\{
\begin{aligned}
  \sum_{i\in p} N_i 
  \mu_c n_{ix} \left(\textbf{u}_p^*-\textbf{u}_c\right)
  \cdot\textbf{n}_i 
  &= 
  \sum_{i\in p} N_i n_{ix} p_c  
  \\
  \sum_{i\in p} N_i 
  \mu_c n_{iy} \left(\textbf{u}_p^*-\textbf{u}_c\right)
  \cdot\textbf{n}_i 
  &=   
  \sum_{i\in p} N_i n_{iy} p_c  
\end{aligned}
\right.
\end{split}
\end{equation}
The velocity components are coupled with each other, thus the inversion of a $2\times 2$ matrix is needed.

The resulting Riemann velocity $\textbf{u}_p^*$ at the node and the Riemann pressure $p_i^*$ at interface segments are used to evaluate the flux on each face of each cell, in addition $\textbf{u}_p^*$ will also be responsible for the mesh motion at vertex.

We note that another form of the shock impedance is $\mu_c = \rho (c + s\delta^*u)$ \cite{burton2013cchSolver}, with $s=\frac{\gamma+1}{2}$ a constant for polytropic gases, and
\begin{equation}
 \delta^*u = 
 \begin{cases}
   \lvert \delta\mathbf{u} \rvert = \lvert \textbf{u}_p^* - \textbf{u}_c \rvert  & \text{compression} \\
   0 & \text{expansion}
 \end{cases}
\end{equation}

It is reported in \cite{burton2015reduction} that using the former acoustic impedance yields the best result, 
despite the second term ($s\delta^*u$) does have a physical basis in approximating the shock velocity.
Although the latter form of the shock impedance will introduce more dissipation than the former acoustic impedance, 
thus potentially resulting in less accurate solution, 
we found it help with the mesh robustness as well as the solution stability in certain cases 
when strong shock presents, for all three solvers considered in this paper.

\subsection{The new LS solver}
As mentioned above, in the new solver proposed in this paper, a single Riemann pressure is needed at a node.
Then the question is, how to determine this single pressure, and also the velocity. 
By observing the Riemann jump equation \ref{eq:RiemannJump}, and noticing that the Riemann pressure and velocity components are the variables whose values are to be determined, we can rewrite the equation as
\begin{equation}
p^* + \mu_c \textbf{u}_p^* \cdot \textbf{n}_i
=
p_c + \textbf{u}_c \cdot \textbf{n}_i
\end{equation} 
where the unknowns are all on the left. Explicitly in 2D, it is
\begin{equation}
p^* + \mu_c n^x_i u_p^* + \mu_c n^y_i v_p^*
=
p_c + \textbf{u}_c \cdot \textbf{n}_i
\end{equation} 
Realizing that this relation is for each interface segment $i$ impinging on node, we rewrite it in the matrix form
\begin{equation}
\begin{pmatrix}
1 & \mu_{c(1)} n^x_1 & \mu_{c(1)} n^y_1\\
1 & \mu_{c(2)} n^x_2 & \mu_{c(2)} n^y_2\\
... & ... & ...\\
1 & \mu_{c(i)} n^x_i & \mu_{c(i)} n^y_i\\
... & ... & ...\\
1 & \mu_{c(N)} n^x_N & \mu_{c(N)} n^y_N
\end{pmatrix}
\begin{pmatrix}
p^*\\
u_p^*\\
v_p^*
\end{pmatrix}
=
\begin{pmatrix}
p_{c(1)} + \textbf{u}_{c(1)} \cdot \textbf{n}_1\\
p_{c(2)} + \textbf{u}_{c(2)} \cdot \textbf{n}_2\\
...\\
p_{c(i)} + \textbf{u}_{c(i)} \cdot \textbf{n}_i\\
...\\
p_{c(N)} + \textbf{u}_{c(N)} \cdot \textbf{n}_N
\end{pmatrix}
=
\begin{pmatrix}
r_1\\
r_2\\
...\\
r_i\\
...\\
r_N
\end{pmatrix}
\end{equation}
The equations above could be solved in the least-squares sense, provided  that the resulting matrix is not singular.
For a typical and valid initial 2D mesh, the number of edges impinging on a node is at least 3, so the number of interface segments $N \geqslant 6$, which indicates that the resulting system is usually overdetermined and a unique solution could be sought.
In 3D, the number of unknowns only increases to 4, in which case the system is still overdetermined.
After all, it is natural to have a unique solution at any point of a physical flow field.  
In very rare cases, the resulting least-squares matrix might be singular.
We will give more detailed discussion in the final manuscript, along with the boundary conditions.

After solving the least-squares problem, the resulting Riemann pressure $p^*$ and velocity $\textbf{u}_p^*$ at the node 
are used to evaluate the numerical flux , in addition $\textbf{u}_p^*$ will also be used for the mesh movement.

We note that in the other two acoustic solvers (by Maire et al.\cite{maire2007siam} and Burton et al. \cite{burton2013cchSolver}), they require information on the interface length (area in 3D).
In the new solver, however, we do not need such information, making this solver rather local,  
i.e., only the interface normals at the node is required, as illustrated in the figure below. 
\begin{figure}[H]
  \centering
  \includegraphics[trim= 5cm 2cm 7cm 1cm,clip,height =2.5in]  
    {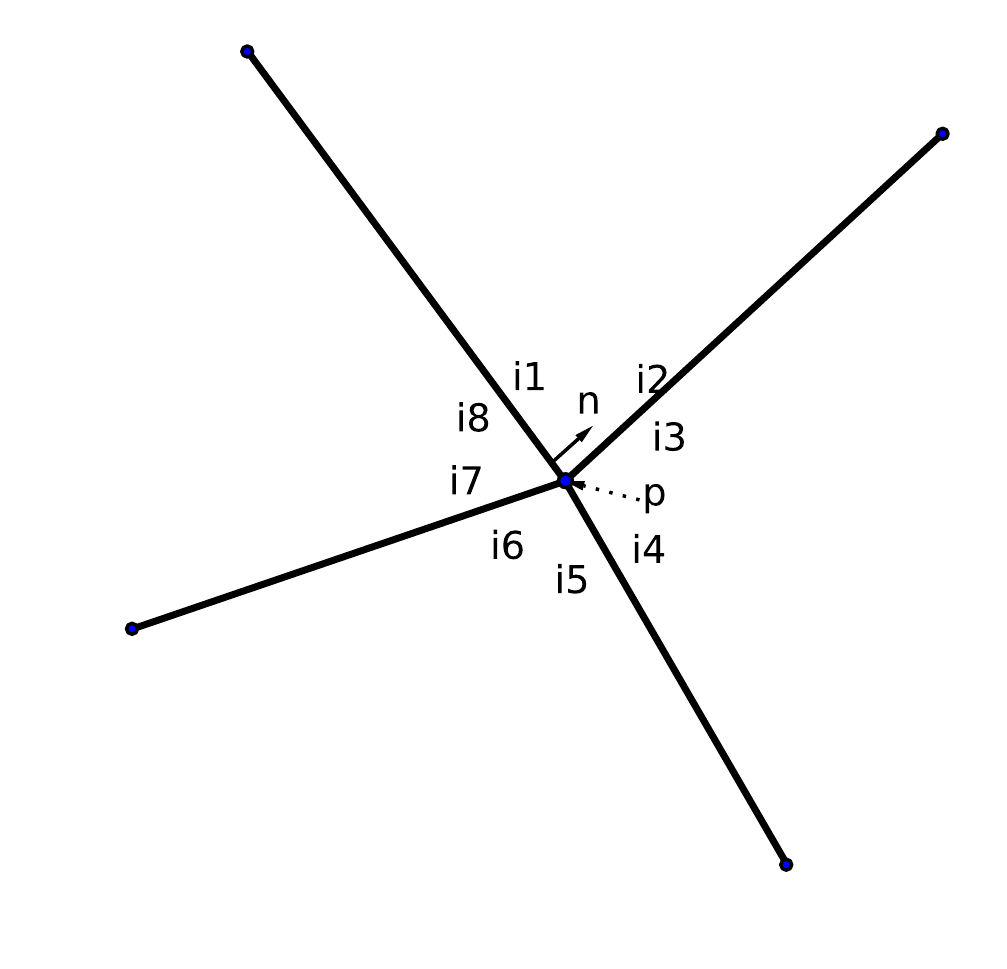}
  \caption{Notations of current LS solver at a node $p$.}
  \label{fig:nodeLS}
\end{figure}

This property might benefit its potential extension to curved elements, since only node local information is used.
\begin{figure}[H]
  \centering
  \includegraphics[trim= 5cm 2cm 7cm 1cm,clip,height =2.5in] 
    {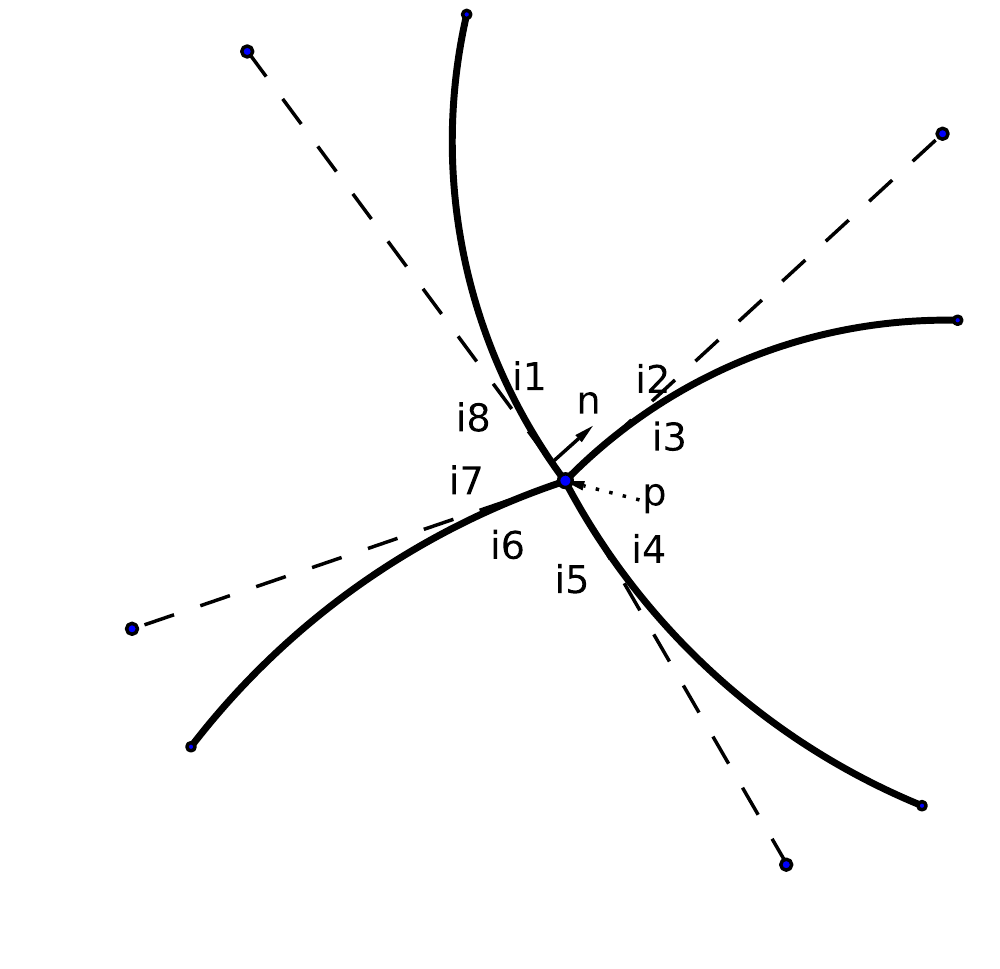}
  \caption{Notations of current LS solver at a node $p$ (curved elements).}
  \label{fig:nodeLS}
\end{figure}

\section{Temporal Discretization}
\label{sec:temporal}

The updated Lagrangian formulation leads to the following semi-discrete system of equations
\begin{equation}
\label{eq:semi-discrete}
\frac{d\left({\Omega\textbf{U}}\right)}{dt}=\textbf{R}
\end{equation}
where $\Omega$ denotes the cell volume, $\textbf{U}$ the global solution vector of the degrees of freedom, and $\textbf{R}$ the right-hand side (RHS).

The time marching for the semi-discrete system Eq. \ref{eq:semi-discrete} can be completed by the Runge-Kutta methods \cite{cockburn1998rk}.
It is worth noting that since the mesh motion is part of the Lagrangian solution, one should use the same temporal discretization for both variable vector evolution and the mesh coordinate trajection. 
In this work, we use a two-stage Runge-Kutta method, and the procedure is shown below:

Given the physical variable vector $\textbf{U}^n$ and the geometrical coordinates $\textbf{x}^n$ at time level $n$, we are seeking the solution vector $\textbf{U}^{n+1}$ and the new locations $\textbf{x}^{n+1}$ for grid point at time level $n+1$, for the time step $\Delta t = t^{n+1} - t^n$.

Step 1: Compute the Riemann velocity $(\textbf{u}^*)^n$ based on $\textbf{U}^n$ and $\textbf{x}^n$.

Step 2: Set the intermediate coordinates $ \textbf{x}^{(1)}$ for each vertex
\[
 \textbf{x}^{(1)} = \textbf{x}^n + \Delta t (\textbf{u}^*)^n 
\]
and compute the intermediate geometrical quantities, such as the cell volume and the face area.

Step 3: Compute the intermediate solution vector $\textbf{U}^{(1)}$ by solving
\[
\Omega^{(1)}\textbf{U}^{(1)} = \Omega^{n}\textbf{U}^{n} + \Delta t \textbf{R}(\textbf{U}^n)
\]

Step 4: Compute the Riemann velocity $(\textbf{u}^*)^{(1)}$ based on $\textbf{U}^{(1)}$ and $\textbf{x}^{(1)}$.

Step 5: Set the $n+1$ level coordinates $ \textbf{x}^{n+1}$ for each vertex
\[
 \textbf{x}^{n+1} = \frac{1}{2}\textbf{x}^n 
 +  \frac{1}{2}\textbf{x}^{(1)} 
 +  \frac{1}{2}\Delta t (\textbf{u}^*)^{(1)} 
\]
and compute the $n+1$ level geometrical quantities.

Step 6: Compute the $n+1$ level solution vector $\textbf{U}^{n+1}$ by solving
\[
\Omega^{n+1}\textbf{U}^{n+1} = 
\frac{1}{2} \Omega^{n}\textbf{U}^{n} 
+\frac{1}{2} \Omega^{(1)}\textbf{U}^{(1)}
+\frac{1}{2} \Delta t \textbf{R}(\textbf{U}^{(1)})
\]

\section{Numerical Examples}
\label{sec:example}
In this section, several numerical test cases have been set up to verify the performance of this new solver, and compare with other two acoustic solvers.
The first two cases will focus on the numerical error and spatial convergence, and the rest ones are to evaluate the solution stability, accuracy and symmetry preservation in the presence of strong shocks.

\subsection{Shockless Noh}
The first test case is the two-dimensional shockless Noh problem \cite{burton2014cchale}.
This is a smooth flow problem with a known analytical solution.
The material is ideal gas with the ratio of specific heats $\gamma = 5/3$.
The initial computational domain is a square $[-1,1]\times[-1,1]$,
with the following initial conditions
\begin{equation}
\begin{aligned}
\rho^0 &= 1 \\
u^0 &= -x^0 \\
v^0 &= -y^0 \\
e^0 &= 1 \\
p^0 &= (\gamma -1)\rho^0 e^0
\end{aligned}
\end{equation}
where $e$ denotes the specific internal energy, $u$ and $v$ the components of the velocity in x- and y- directions.
A Dirichlet boundary condition for the velocity is prescribed during the simulation, with no dependence on time, 
i.e., for any boundary node $(x_b, y_b)$, a constant velocity from the initial condition and its initial position $(x_b^0, y_b^0)$ is attached to it,
\begin{equation}
\begin{aligned}
&u(x_b,y_b,t) = u(x_b^0, y_b^0,0) = -x_b^0 \\
&v(x_b,y_b,t) = v(x_b^0, y_b^0,0) = -y_b^0
\end{aligned}
\end{equation}

\begin{figure}[H]
  \centering
  \subfloat[initial mesh]{ 
    \label{fig:smoothNoh:mesh:initial} 
    \includegraphics[trim= 2cm 1.5cm 2.48cm 0.5cm,clip,height = 2in]
    {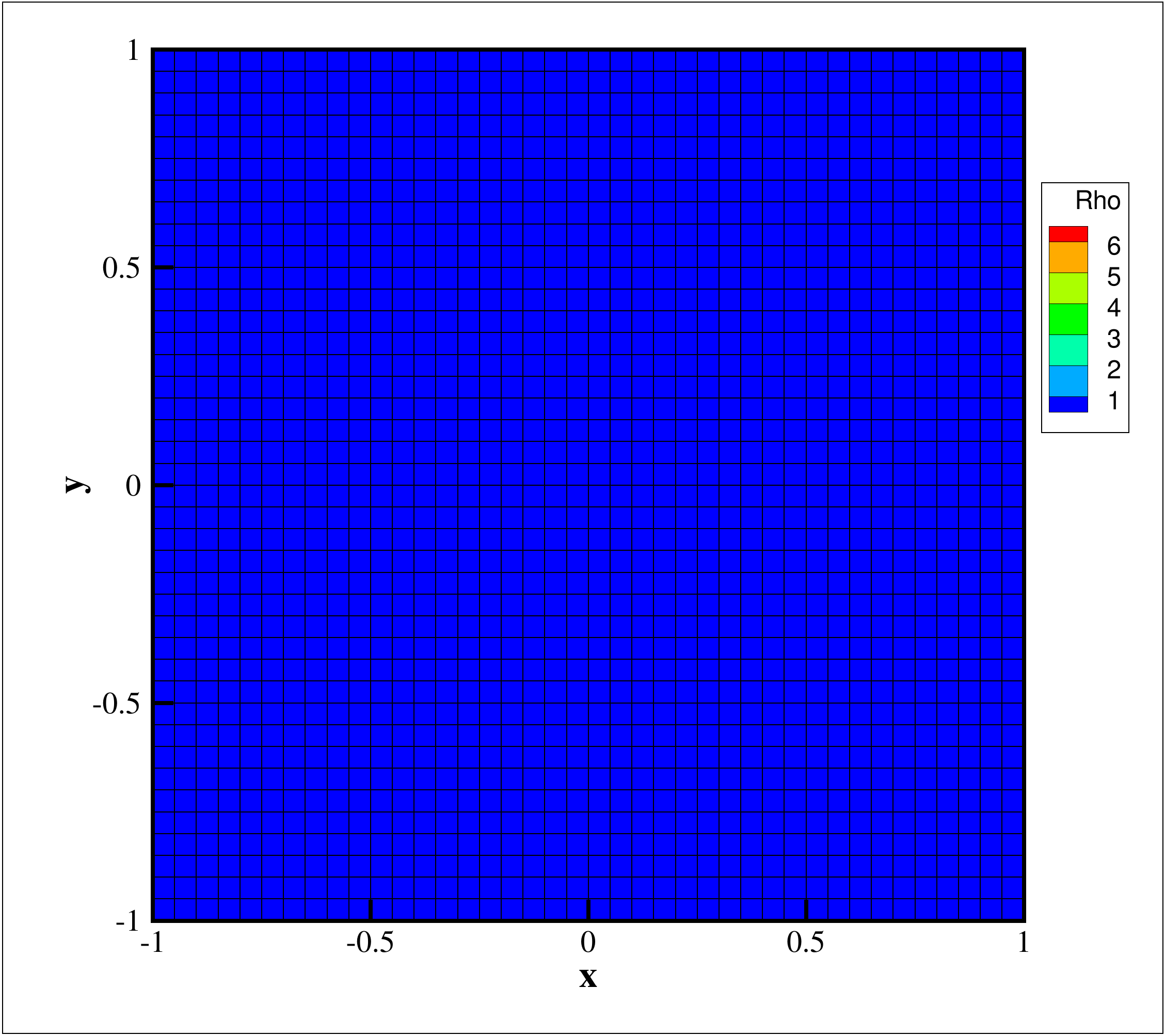} 
  }
  \subfloat[mesh at $t = 0.6$]{ 
    \label{fig:smoothNoh:mesh:final}   
    \begin{minipage}[t][2in][t]{0.5\linewidth}
      \centering
      \includegraphics[trim= 2cm 1.5cm 2.48cm 0.5cm,clip,height = 2in]
      {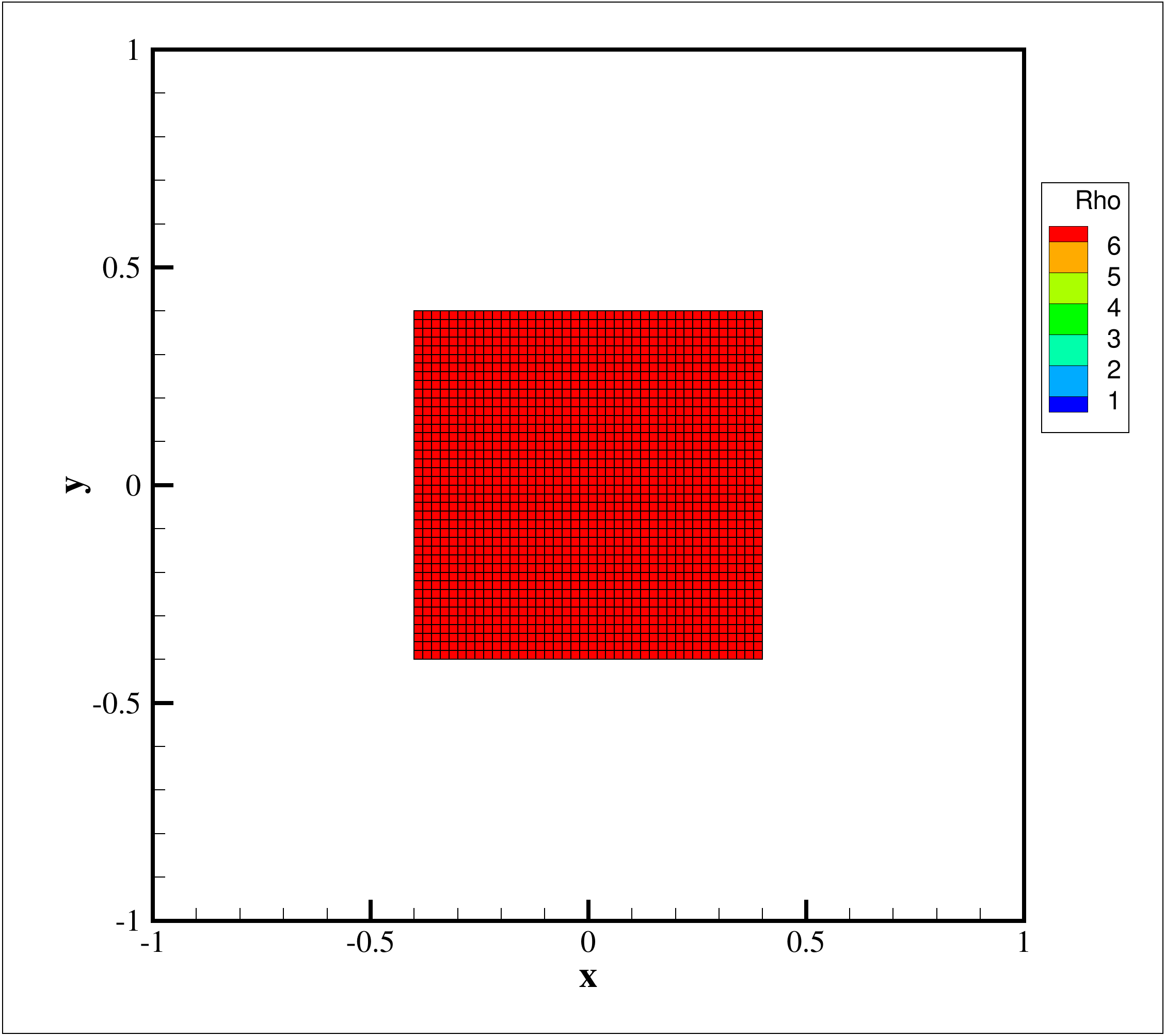}
      \includegraphics[trim= 20.3cm 3.5cm 0.5cm 0.5cm,clip,height = 2in]
      {./fig/smoothNoh/final-mesh-eps-converted-to.pdf}
    \end{minipage}
  }
  \caption{Initial and final grids of the shockless Noh problem, contoured by density.}
  \label{fig:smoothNoh:mesh}
\end{figure}

The analytical solution of this problem is dictated by
\begin{equation}
\begin{aligned}
&\rho(x,y,t) = \rho^0\left(1-t \right)^{-\alpha} \\
&u(x,y,t) = u^0 \\
&v(x,y,t) = v^0 \\
&e(x,y,t) = e^0\left(1-t \right)^{-\alpha\left(\gamma-1 \right)}
\end{aligned}
\end{equation}
with $\alpha=2$. 
It can be seen that the density $\rho$ and specific internal energy $e$ are spatially invariant, and are only functions of time.
We use this test case to assess the spatial convergence.
The mesh refinement involves a set of five uniform grids with quadrilateral elements: $10\times 10$, $20\times 20$, $40\times 40$, $80\times 80$ and $160\times 160$. 
The initial mesh and density distribution are illustrated in Fig. \ref{fig:smoothNoh:mesh:initial}. 
The simulation stops at $t=0.6$ and the final mesh and density contour are shown in Fig. \ref{fig:smoothNoh:mesh:final}.
  
To assess the order of accuracy, we compute the $L_2$ norm of the numerical error from internal energy at $t=0.6$.
Table \ref{tab:smoothNoh:l2norm} shows the numerical error obtained on the set of five grids using the three nodal Riemann solvers.

\begin{table}[ht]
\begin{center}
\caption{Spatial accuracy and convergence rate for the shockless Noh problem at $t = 0.6$.}
\begin{tabular}{r|cc|cc|cc}
\hline
\multirow{2}{*}{Mesh} &  \multicolumn{2}{c|}{Burton solver}  &  \multicolumn{2}{c|}{Maire solver} &  \multicolumn{2}{c}{LS solver} \\
\cline{2-7}
 & $L_2$ error & order & $L_2$ error & order & $L_2$ error & order \\
\hline
10 $\times$ 10   & 5.9324E-1 &  -     & 4.1626E-1 &  -    & 4.1626E-1 &  -     \\
20 $\times$ 20   & 2.8692E-1 &  1.05  & 2.0208E-1 &  1.04 & 2.0208E-1 &  1.04  \\
40 $\times$ 40   & 1.4108E-1 &  1.02  & 9.9556E-2 &  1.02 & 9.9556E-2 &  1.02  \\
80 $\times$ 80   & 6.9949E-2 &  1.01  & 4.9411E-2 &  1.01 & 4.9411E-2 &  1.01  \\
160 $\times$ 160 & 3.4828E-2 &  1.01  & 2.4614E-2 &  1.00 & 2.4614E-2 &  1.00  \\
\hline
\end{tabular}
\label{tab:smoothNoh:l2norm}
\end{center}
\end{table}
One can see that all three solutions achieve the designed 1st order convergence.
For this special case, the new LS solver leads to the same numerical error as the Maire solver; and both errors are smaller than that from the Burton solver.

\subsection{Taylor-Green Vortex}

\begin{figure}[H]
  \centering
  \subfloat[initial mesh]{
    \label{fig:tgv:mesh:initial}  
    \includegraphics[trim= 2cm 1.5cm 1.0cm 0.5cm,clip,height = 2in]
    {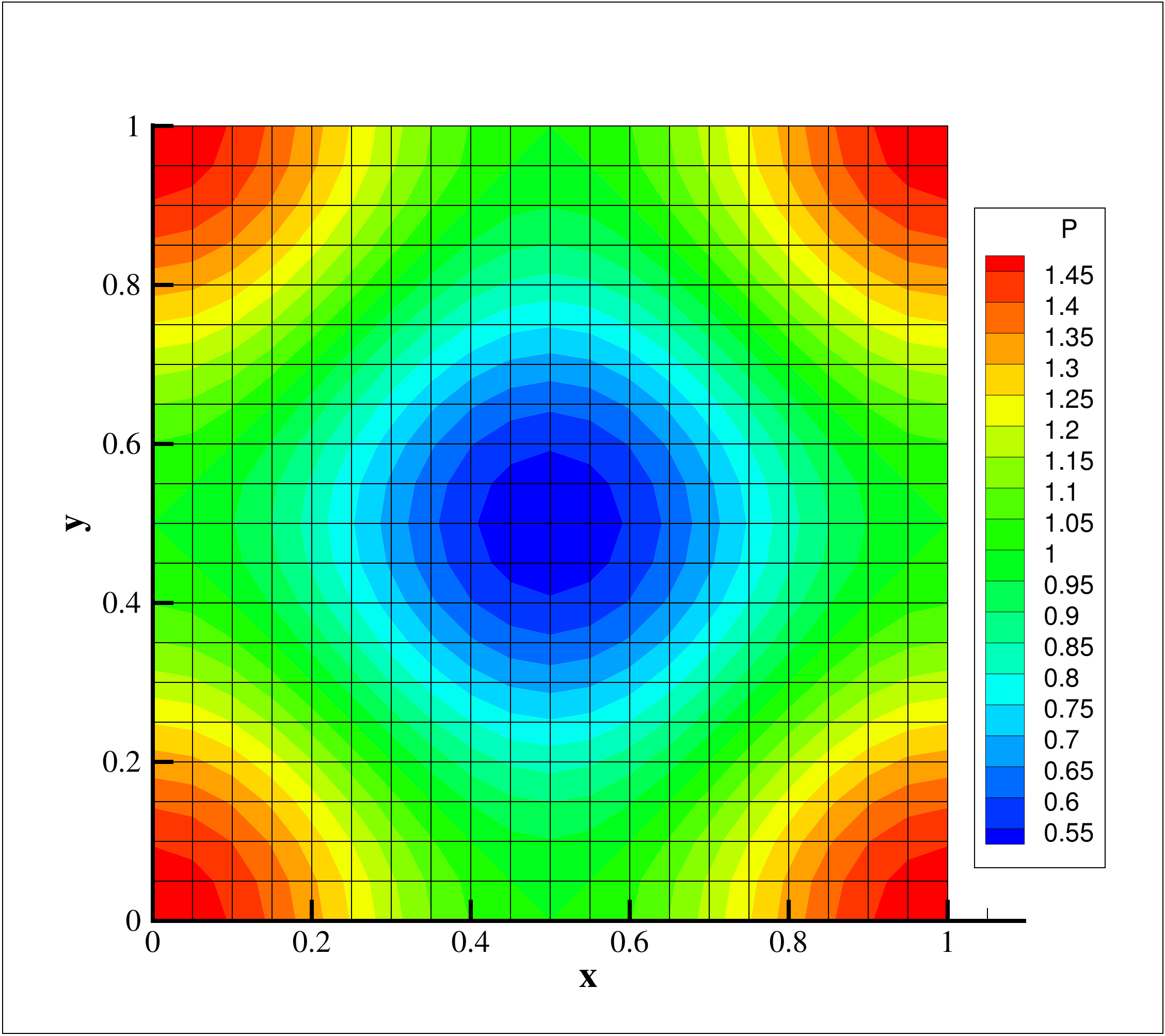} }
  \subfloat[mesh at $t = 0.4$]{
    \label{fig:tgv:mesh:final}  
    \includegraphics[trim= 2cm 1.5cm 1.0cm 0.5cm,clip,height = 2in]
    {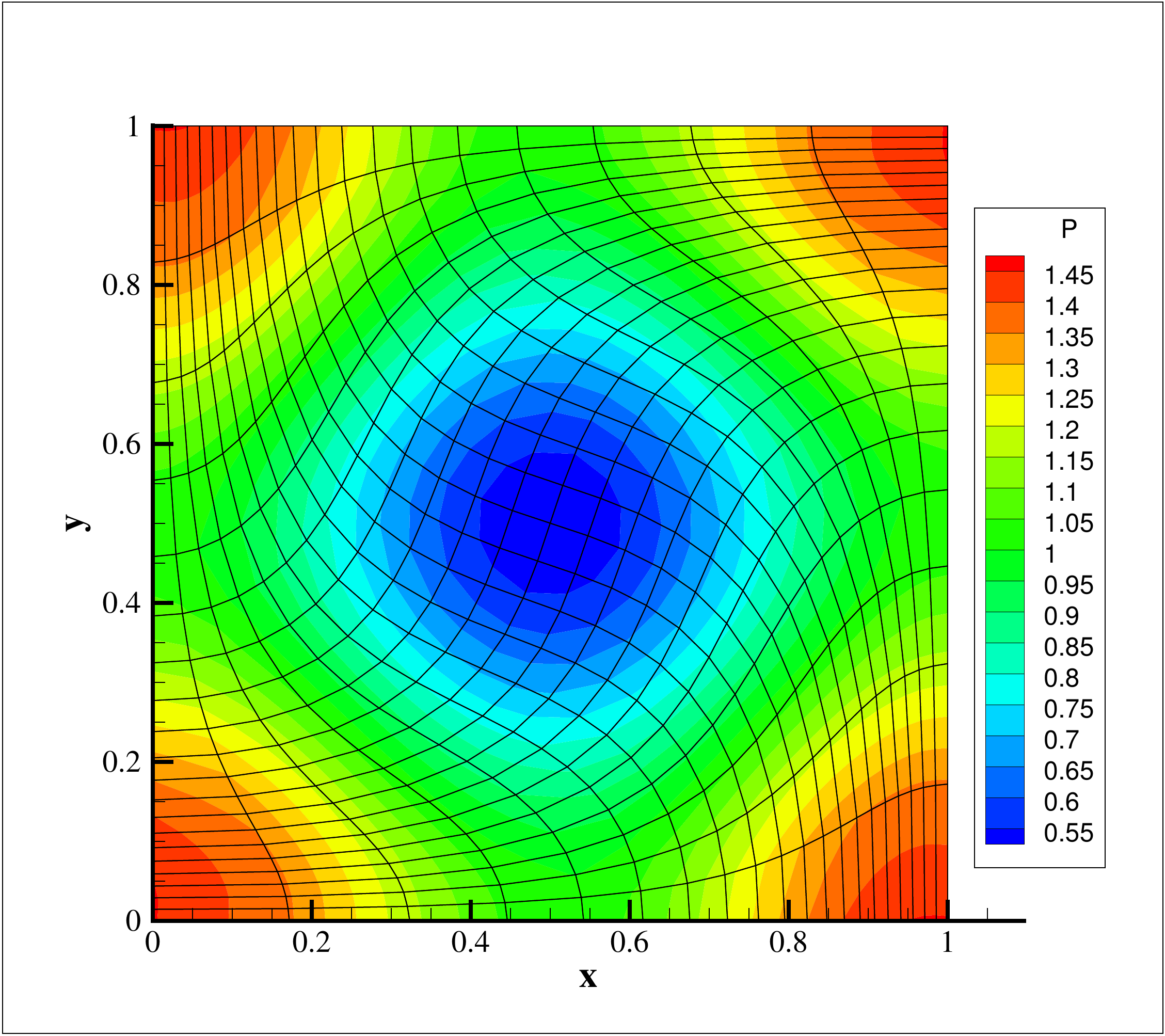} }
  \caption{Initial and final grids of the Taylor-Green vortex problem.}
  \label{fig:tgv:mesh}
\end{figure}

The 2D Taylor-Green vortex problem \cite{dobrev2012tgv,vilar2012ccdg,
burton2015reduction,morgan2015godunovpch} is another benchmark test case with analytical smooth solution thus permitting convergence analysis.
The initial condition are prescribed by
\begin{equation}
\begin{aligned}
&\rho^0 = 1 
\\
&u^0 = \text{sin}\left(\pi x\right)\text{cos}\left(\pi y\right) 
\\
&v^0 = -\text{cos}\left(\pi x\right)\text{sin}\left(\pi y\right) 
\\
&p^0 = \frac{1}{4}\left[\text{cos}\left(2\pi x\right)
+ \text{cos}\left(2\pi y\right) 
\right] + 1 
\\
&e^0 = \frac{p^0}{\rho^0\left(\gamma -1 \right)} 
+ \frac{1}{2}\left( \left(u^0 \right)^2 + \left(v^0 \right)^2\right)
\end{aligned}
\end{equation}  
where $e$ denotes the specific internal energy, $u$ and $v$ the velocity components in the x- and y- directions, respectively.
The flow material is ideal gas with $\gamma = 7/5$.
We note that the above definition of density, velocity and pressure implies that the continuity and momentum equations are automatically satisfied. 
However, to make the flow steady state, a source term in the energy equation is required
\begin{equation}
S = \frac{\pi}{4\left(\gamma -1\right)}\left[ 
\text{cos}\left(3\pi x\right)\text{cos}\left(\pi y\right)
- \text{cos}\left(\pi x\right)\text{cos}\left(3\pi y\right)
\right]
\end{equation}
The computational domain is a square $[0,1]\times [0,1]$, consisting of uniform quadrilateral grids, as is shown in Fig. \ref{fig:tgv:mesh:initial} together with the initial pressure distribution.
The simulation is carried out until $t=0.4$, and the final mesh and pressure contour are shown in Fig. \ref{fig:tgv:mesh:final}.
 
In order to evaluate the spatial convergence rate, a mesh refinement study is performed on successively refined grids: $10\times 10$, $20\times 20$, $40\times 40$, $80\times 80$ and $160\times 160$.
The $L_2$ numerical errors of the pressure are computed at the end time $t=0.4$ and listed in Table \ref{tab:tgv:l2norm}.
We can see that the new LS solver delivers smaller absolute error and higher convergence rate, than the other two solvers.
Nevertheless, all three solvers are approaching 1st order convergence  when refining the mesh further.

\begin{table}[ht]
\begin{center}
\caption{Spatial accuracy and convergence rate for Taylor-Green vortex problem at $t = 0.4$.}
\begin{tabular}{r|cc|cc|cc}
\hline
\multirow{2}{*}{Mesh}  &  \multicolumn{2}{c|}{Burton solver}  &  \multicolumn{2}{c|}{Maire solver} &  \multicolumn{2}{c}{LS solver} \\
\cline{2-7}
 & $L_2$ error & order & $L_2$ error & order & $L_2$ error & order \\
\hline
10 $\times$ 10   & 1.4627E-1 &  -    & 1.1531E-1 &  -     & 8.6650E-2 & - \\
20 $\times$ 20   & 8.2653E-2 &  0.82 & 6.8966E-2 &  0.74  & 5.5295E-2 & 0.65 \\
40 $\times$ 40   & 4.4976E-2 &  0.88 & 4.1131E-2 &  0.75  & 3.1038E-2 & 0.83 \\
80 $\times$ 80   & 2.4200E-2 &  0.89 & 2.3415E-2 &  0.81  & 1.6415E-2 & 0.92 \\
160 $\times$ 160 & 1.2842E-2 &  0.91 & 1.2737E-2 &  0.88  & 8.4503E-3 & 0.96 \\
\hline
\end{tabular}
\label{tab:tgv:l2norm}
\end{center}
\end{table}

\subsection{Sod Shock Tube}

The Sod shock tube problem \cite{sod1978} is a classical test case for validating and verifying numerical schemes.
The specific heats ratio of the ideal gas is $\gamma = 7/5$.
In \cite{vilar2012ccdg}, the polar geometry and mesh were used for this problem, as an extension to the original Cartesian one.
In this work, we consider both the 1D Cartesian case and the 2D polar case.

\subsubsection{1D Cartesian case}
To set up this 1D problem, a computational domain of $[0,1]\times[0,1]$ is selected, with 100 uniform elements in the x- direction and 2 cell-layers in the y-direction.
The contact discontinuity is located at $x=0.5$ at the initial time.
To the left and right are two uniform states in space.
The left state with a high pressure is given as $(\rho^0,u^0,v^0,p^0)_L = (1,0,0,1)$, and the right state is prescribed by $(\rho^0,u^0,v^0,p^0)_R = (0.125,0,0,0.1)$.
The computation is run up to time $t = 0.2$.
The computed solutions for three solvers are plotted in the figures below.
\begin{figure}[H]
  \centering
  \includegraphics[height =2.5in]
    {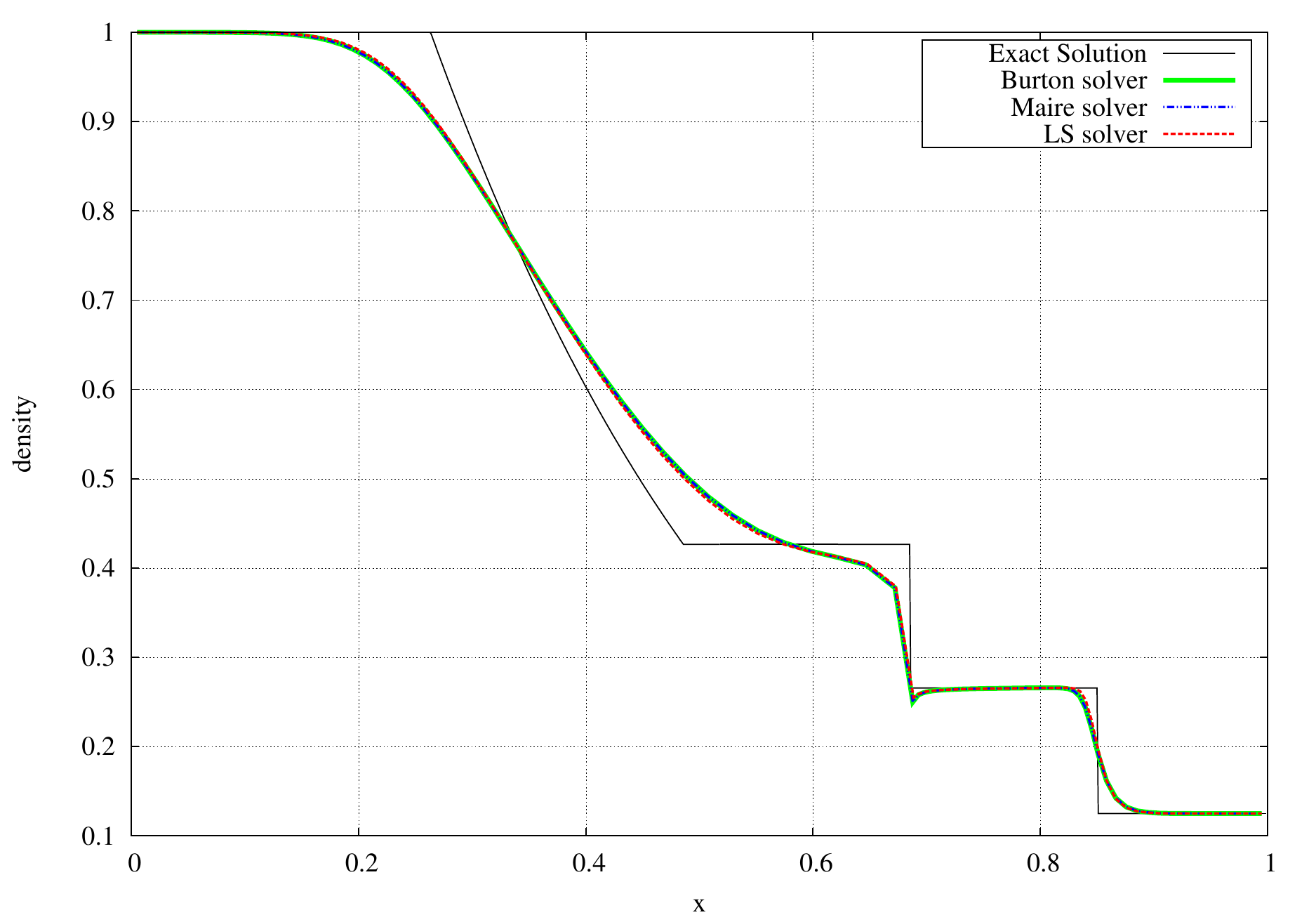}
  \caption{Density profile for the 1D Cartesian Sod shock tube problem.}
  \label{fig:sod:density}
\end{figure}

\begin{figure}[H]
  \centering
  \includegraphics[height =2.5in]
    {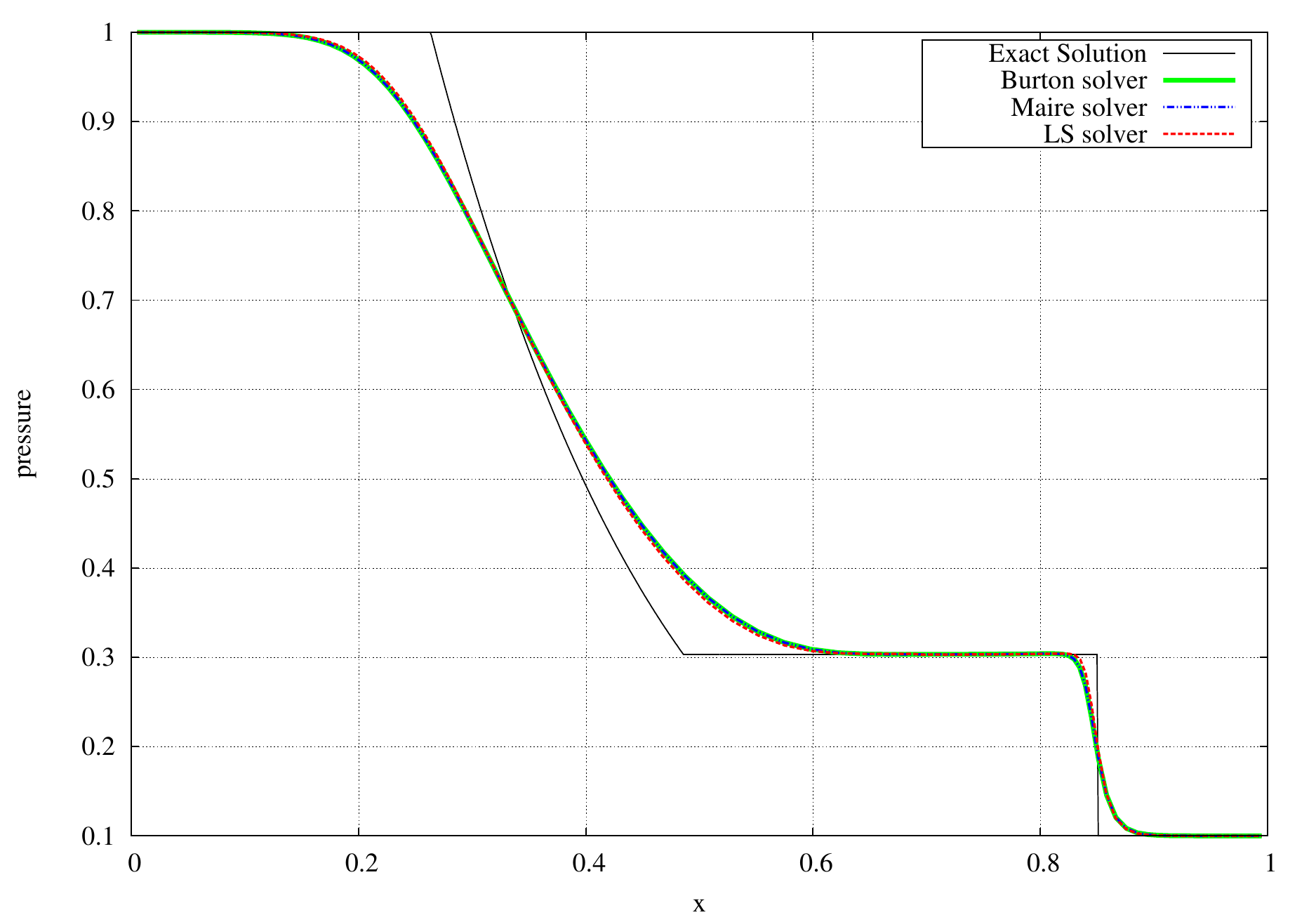}
  \caption{Pressure profile for the 1D Cartesian Sod shock tube problem.}
  \label{fig:sod:pressure}
\end{figure}

\begin{figure}[H]
  \centering
  \includegraphics[height =2.5in]
    {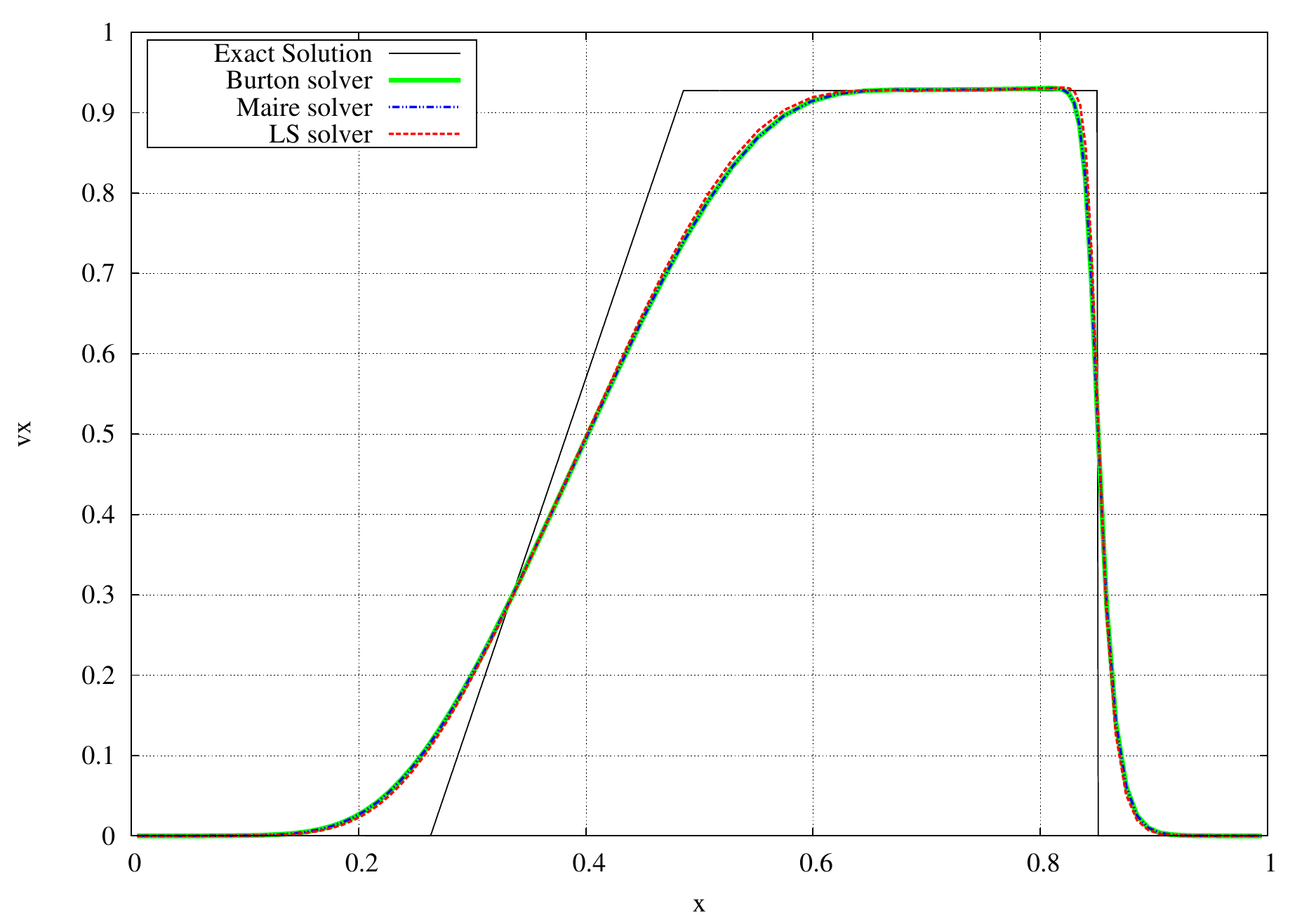}
  \caption{Velocity profile for the 1D Cartesian Sod shock tube problem.}
  \label{fig:sod:vx}
\end{figure}

We can see from the above density, pressure and velocity plots that these three solvers result in almost the same solutions.
If we zoom in, the new LS solver shows slightly better performance. 

\begin{figure}[H]
  \centering
  \includegraphics[height =2.5in]
    {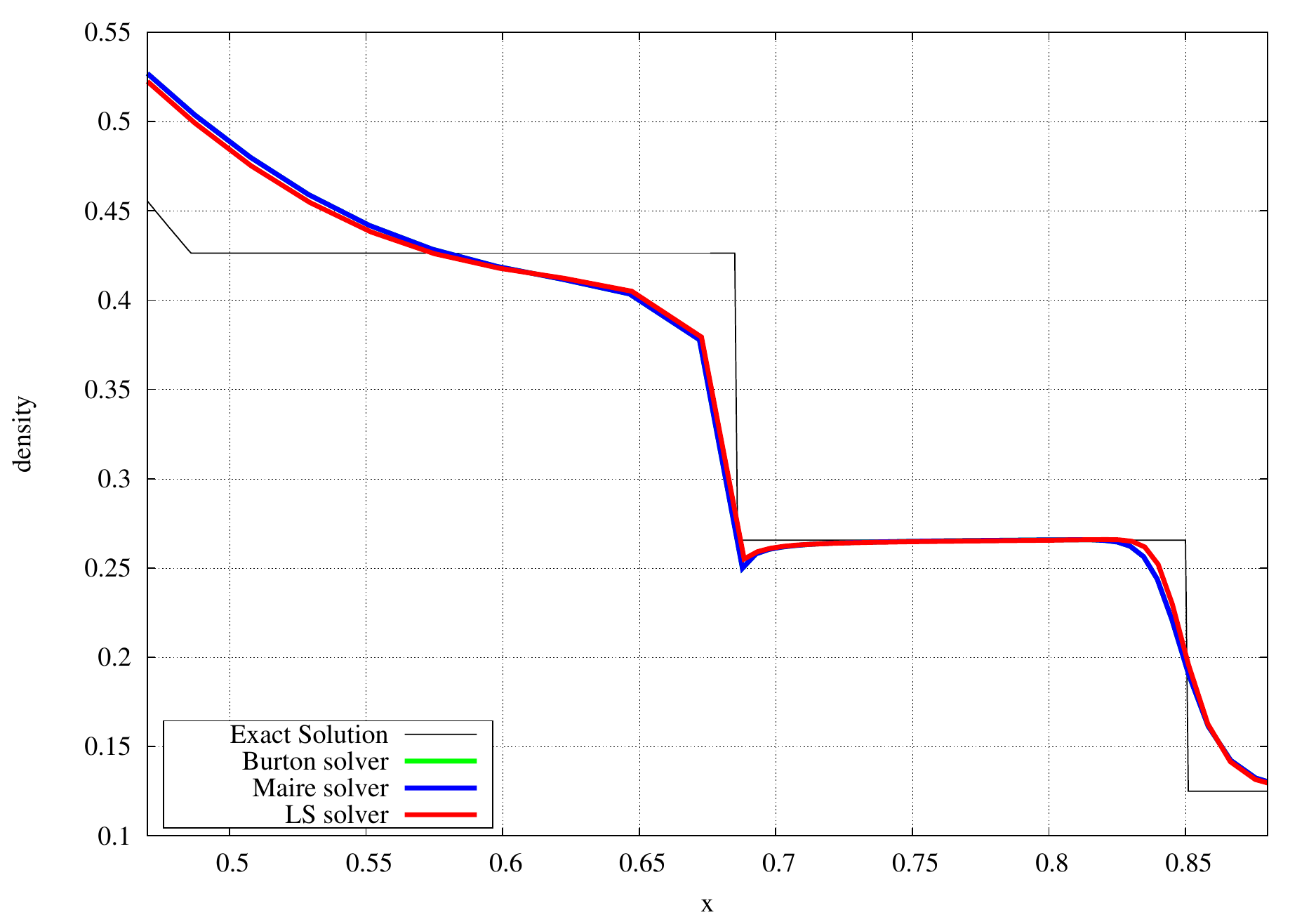}
  \caption{Density profile (zoomed-in) for the 1D Cartesian Sod shock tube problem.}
  \label{fig:sod:density_zoomin}
\end{figure}

\begin{figure}[H]
  \centering
  \includegraphics[height =2.5in]
    {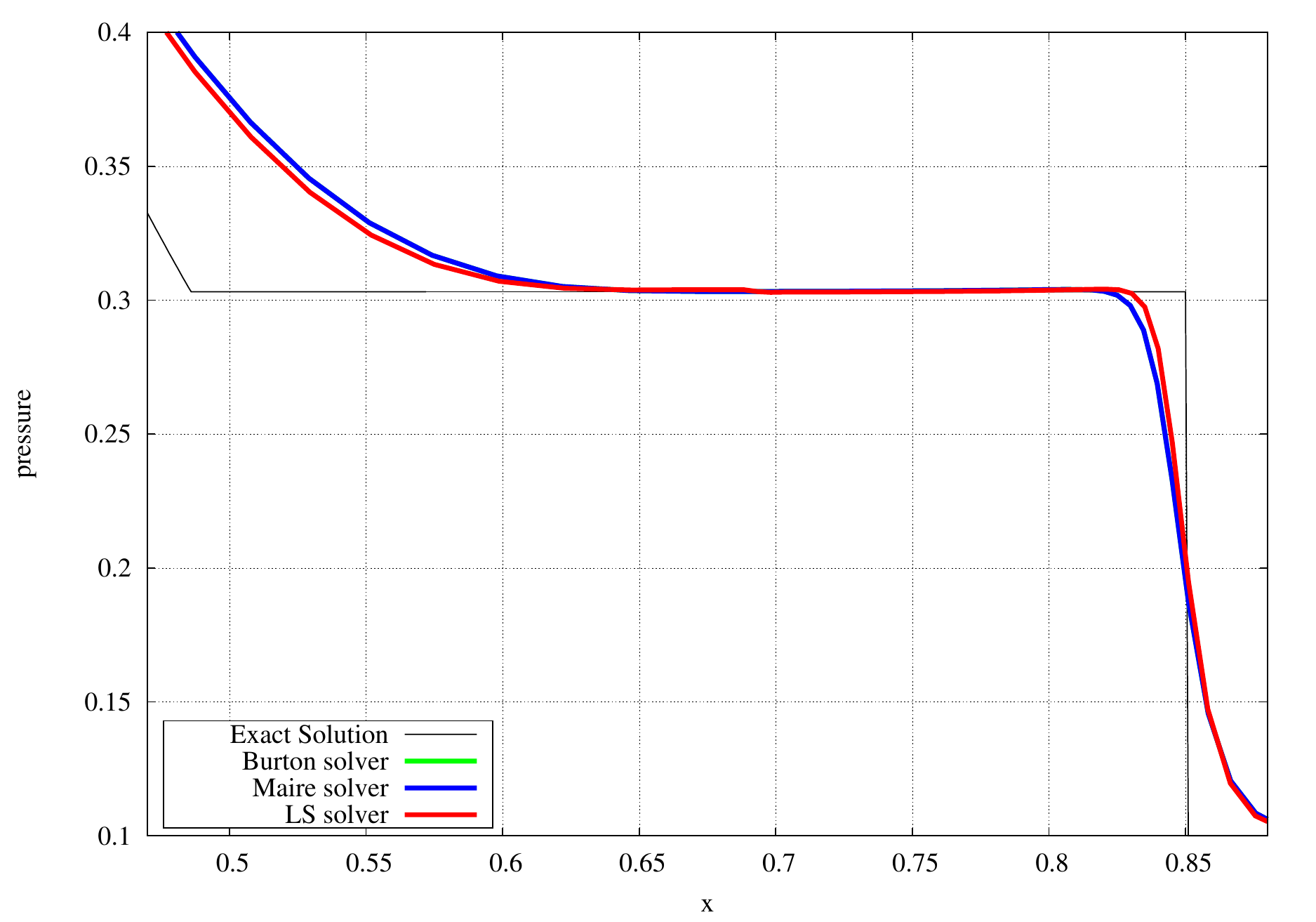}
  \caption{Pressure profile (zoomed-in) for the 1D Cartesian Sod shock tube problem.}
  \label{fig:sod:pressure_zoomin}
\end{figure}

\begin{figure}[H]
  \centering
  \includegraphics[height =2.5in]
    {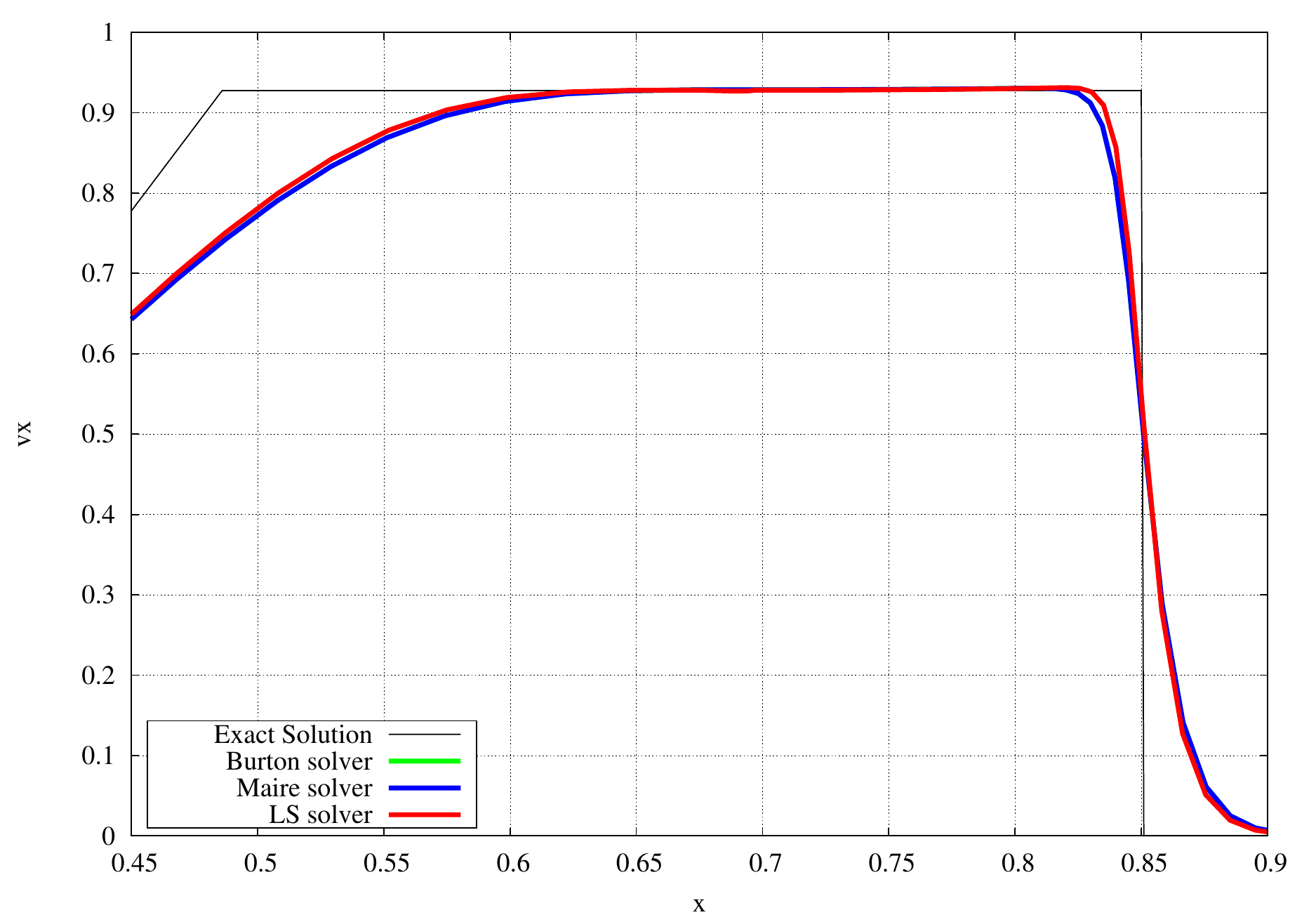}
  \caption{Velocity profile (zoomed-in) for the 1D Cartesian Sod shock tube problem.}
  \label{fig:sod:vx_zoomin}
\end{figure}

\subsubsection{2D polar case}

\begin{figure}[H]
  \centering
  \subfloat[initial mesh]{
    \includegraphics[trim= 1.0cm 0.7cm 2.0cm 1.5cm,clip,height = 2.5in]
    {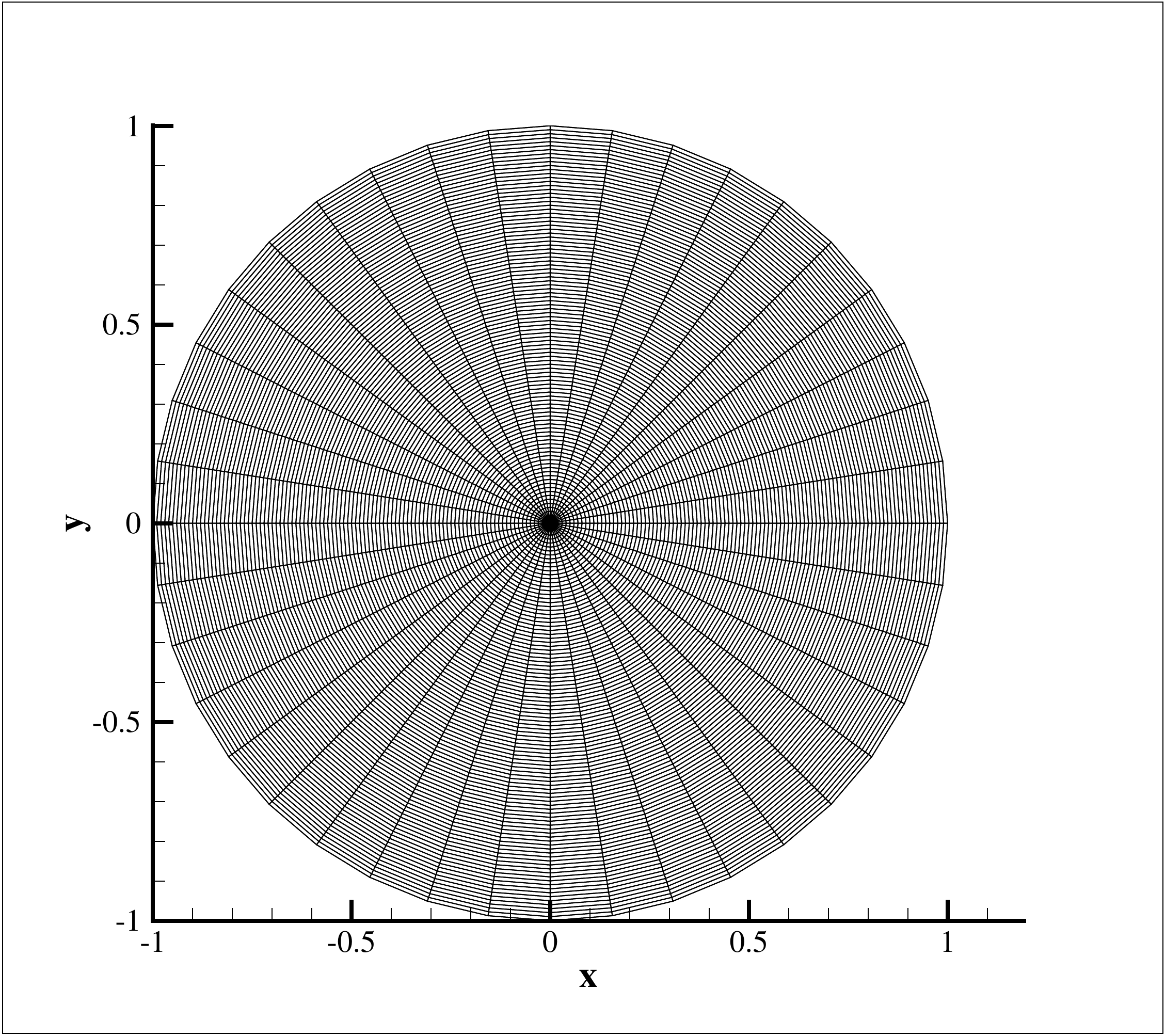} }
  \subfloat[mesh near origin]{
    \includegraphics[trim= 1.0cm 0.7cm 1.0cm 0.5cm,clip,height = 2.5in]
    {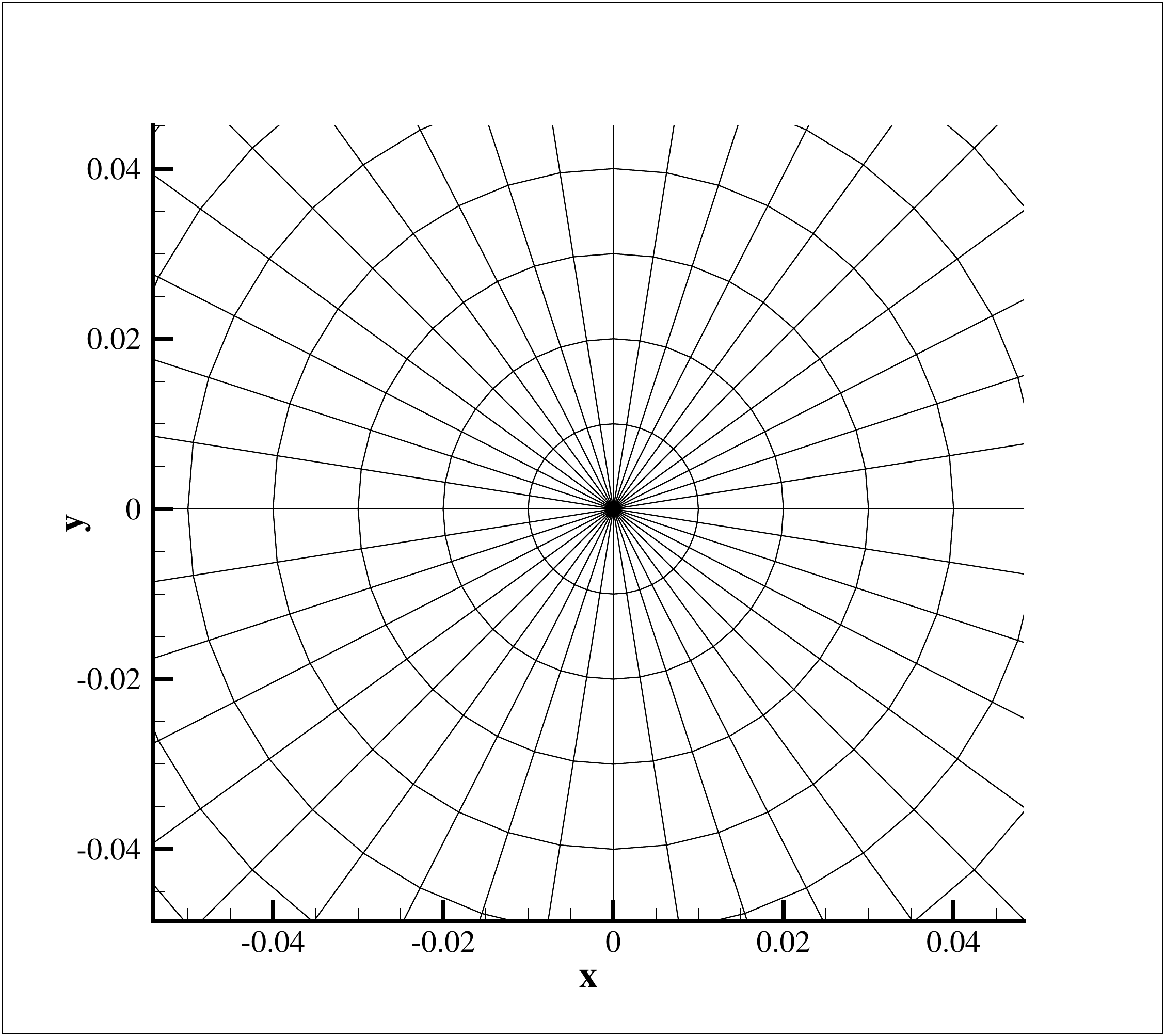} }
  \caption{Initial grids for the 2D polar Sod shock tube problem.}
  \label{fig:polarSod:mesh:initial}
\end{figure}

For the polar case, we consider the computational domain of a circular region with unit radius on $[0,1]\times[0,2\pi]$.
The mesh size is $100\times40$, i.e., 100 elements are placed in the radial direction, and 10 elements in the circumferential direction for each of the four quadrants, as shown in Fig. \ref{fig:polarSod:mesh:initial}.
The contact discontinuity is initially located at $r=0.5$, 
and the left state and right state are given as $(\rho^0,u^0,v^0,p^0)_L = (1,0,0,1)$ and $(\rho^0,u^0,v^0,p^0)_R = (0.125,0,0,0.1)$, respectively, with $u$ and $v$ the x- and y- velocity components.
The simulation is run to time $t = 0.2$.

\begin{figure}[H]
  \centering
  \subfloat[final mesh]{
    \includegraphics[trim= 2cm 1.5cm 1.0cm 1.5cm,clip,height = 2.5in]
    {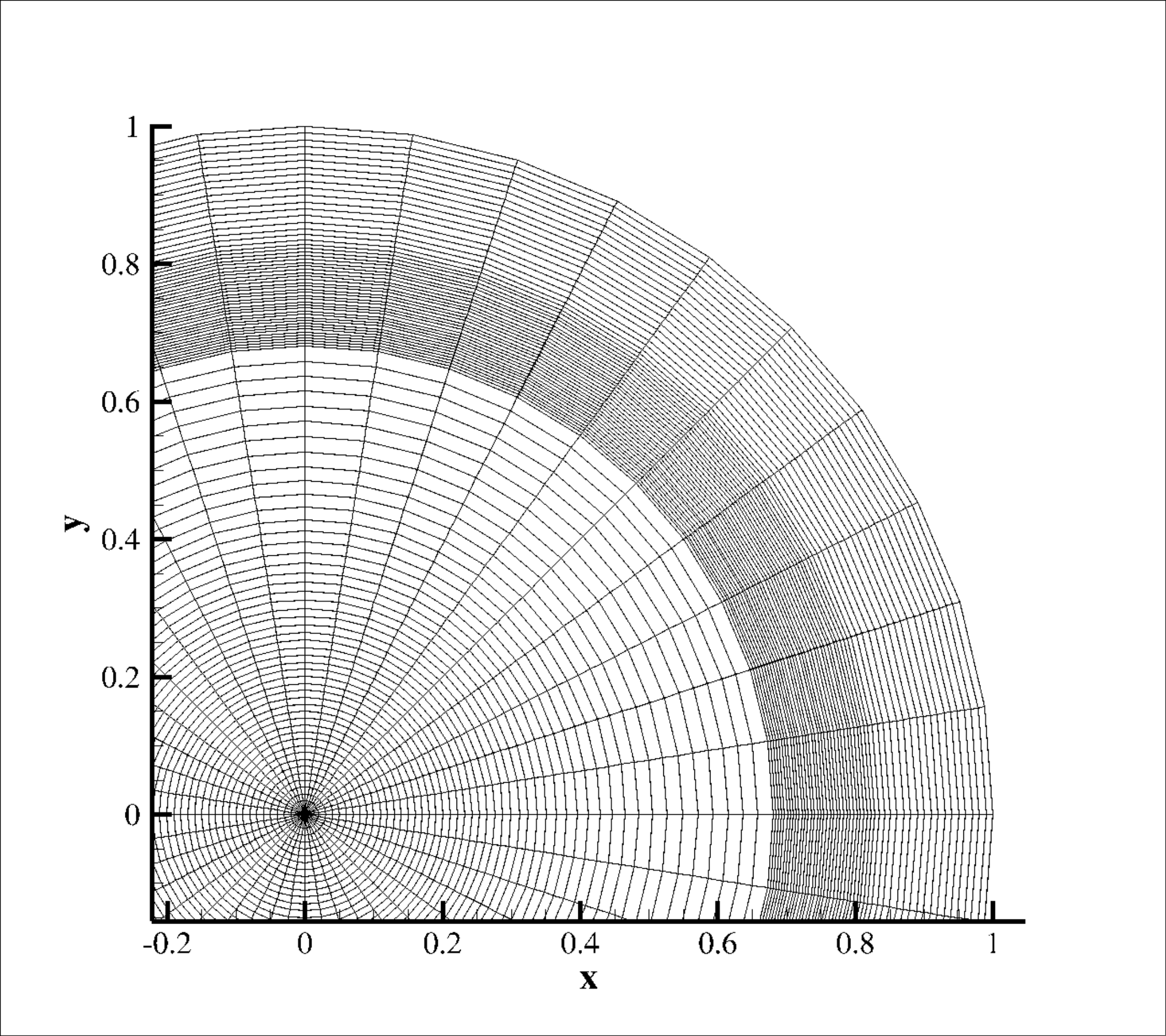} }
  \subfloat[final density]{
    \includegraphics[trim= 2cm 1.5cm 1.0cm 1.5cm,clip,height = 2.5in]
    {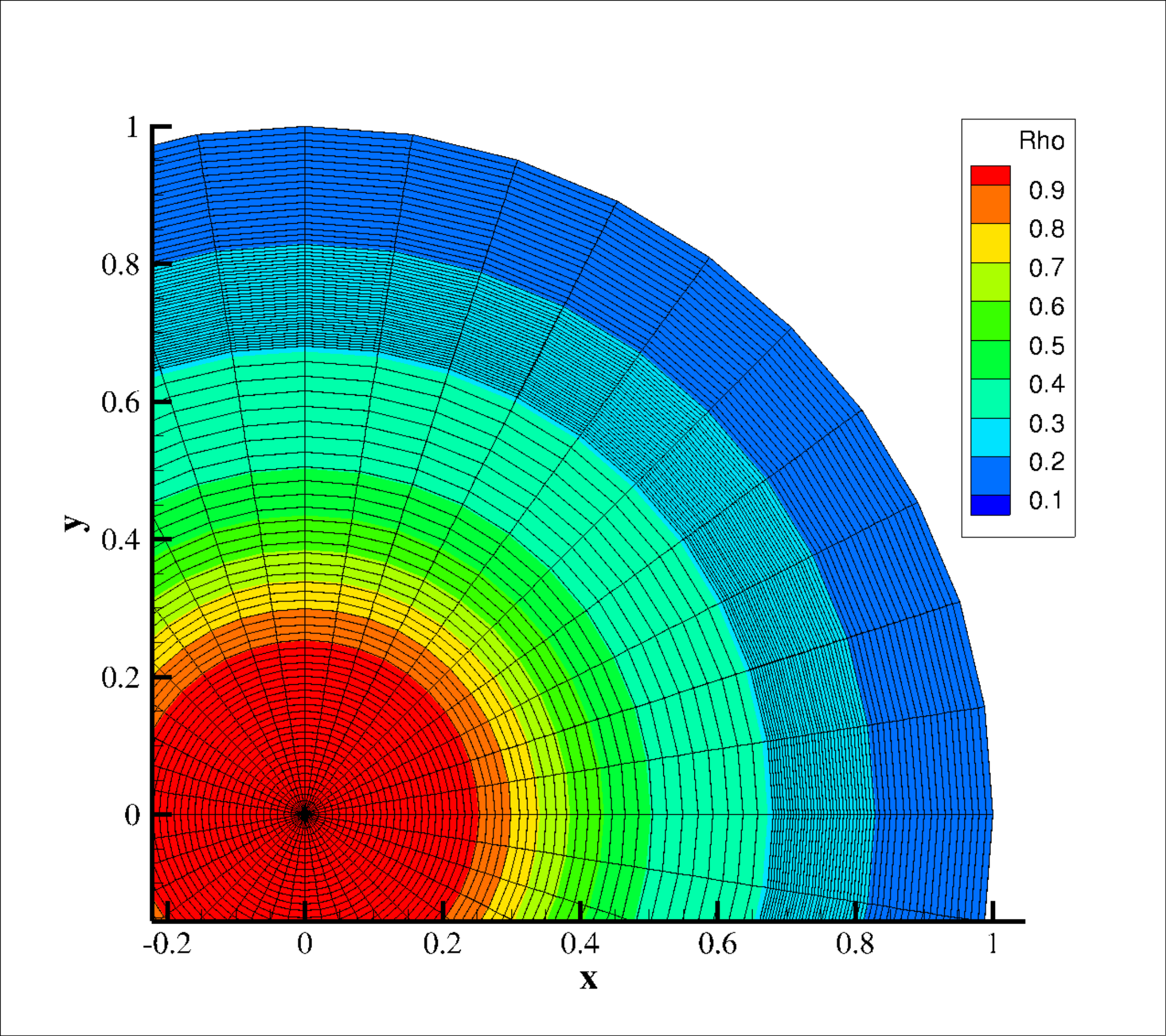} }
  \caption{Final mesh and density distribution for the 2D polar Sod shock tube problem.}
  \label{fig:polarSod:mesh-density:final}
\end{figure}

The final mesh and density contour are displayed in Fig. \ref{fig:polarSod:mesh-density:final}.
The contact discontinuity and shock position can be seen clearly in the mesh.
The results are observed to be quite axisymmetric.

\begin{figure}[H]
  \centering
  \includegraphics[height =2.5in]
    {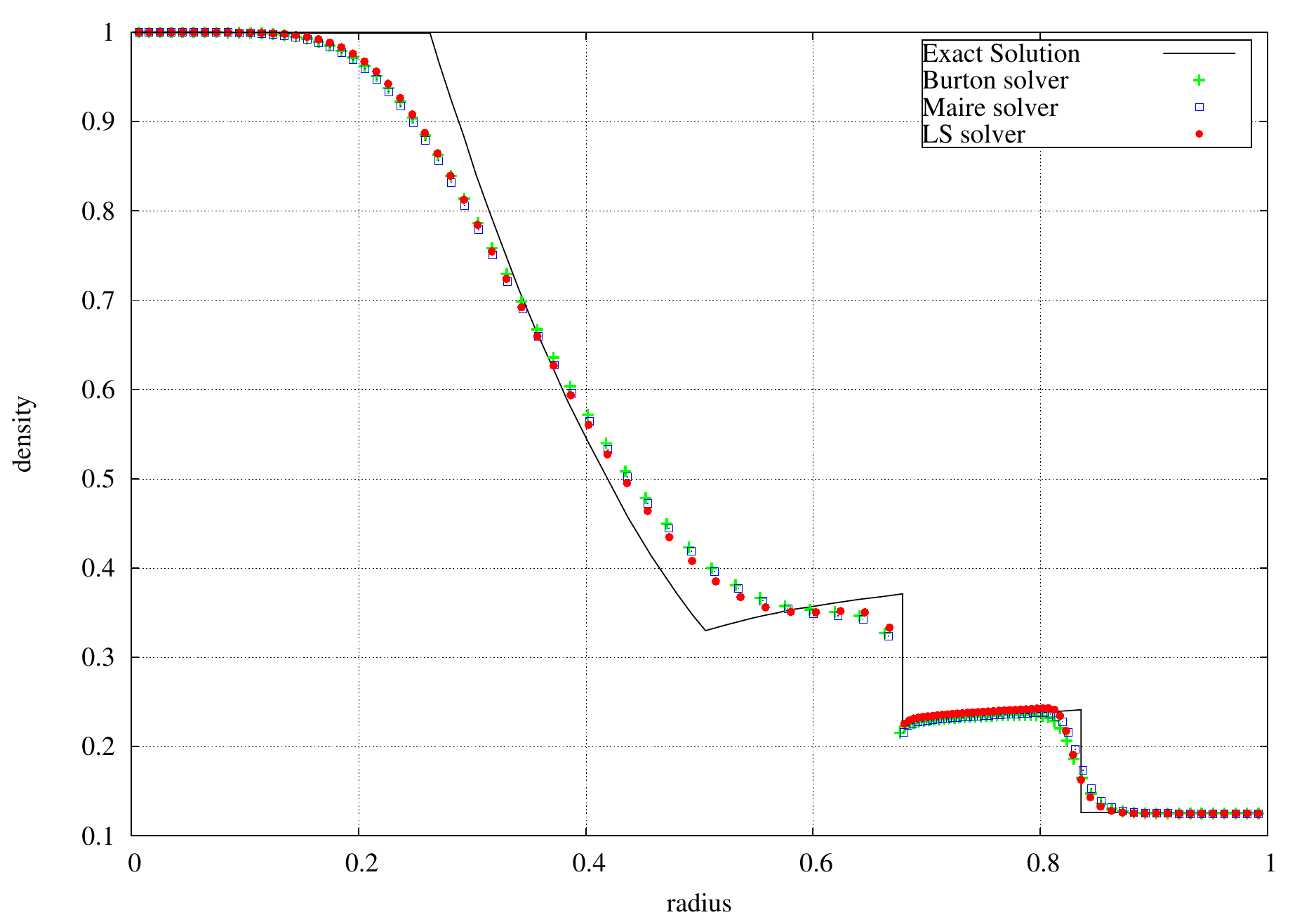}
  \caption{Density profile for the 1D Cartesian Sod shock tube problem.}
  \label{fig:polarSod:plot:density}
\end{figure}

\begin{figure}[H]
  \centering
  \includegraphics[height =2.5in]
    {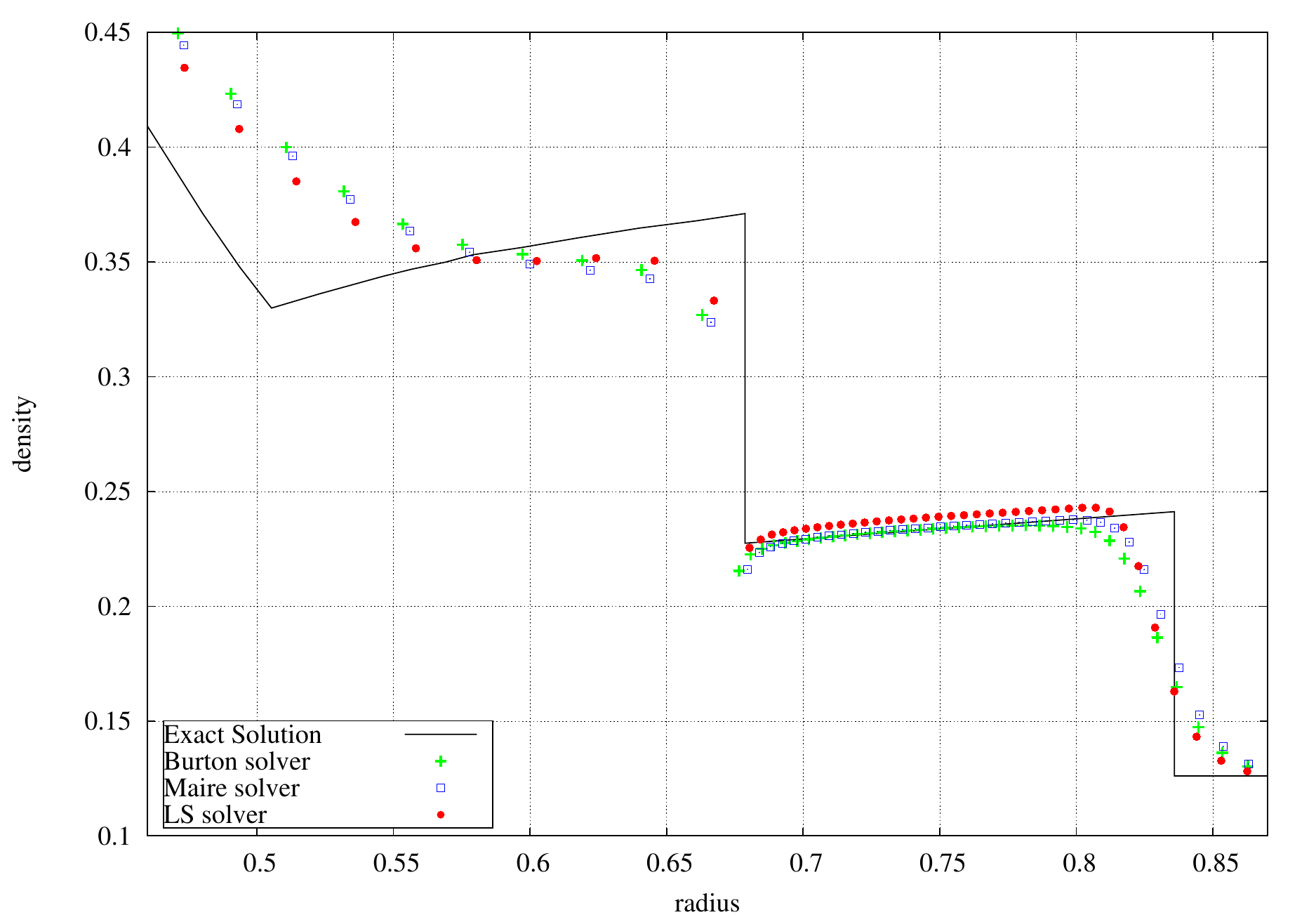}
  \caption{Density profile (zoomed-in) for the 2D polar Sod shock tube problem.}
  \label{fig:polarSod:plot:density_zoomin}
\end{figure}

As in the 1D Cartesian case, we proceed by creating the scattered plots of the density at the cell-center of each element in the mesh, versus the radius of the corresponding cell-center coordinates.
The full profiles are shown in Fig. \ref{fig:polarSod:plot:density}, and the zoomed-in ones are given in Fig. \ref{fig:polarSod:plot:density_zoomin}.
Again, the difference between the new LS solver and the other two is quite slight.

\subsection{Sedov Blast Wave}

\begin{figure}[H]
  \centering 
  \includegraphics[trim= 2cm 1.5cm 1.0cm 0.5cm,clip,height = 2.0in]
  {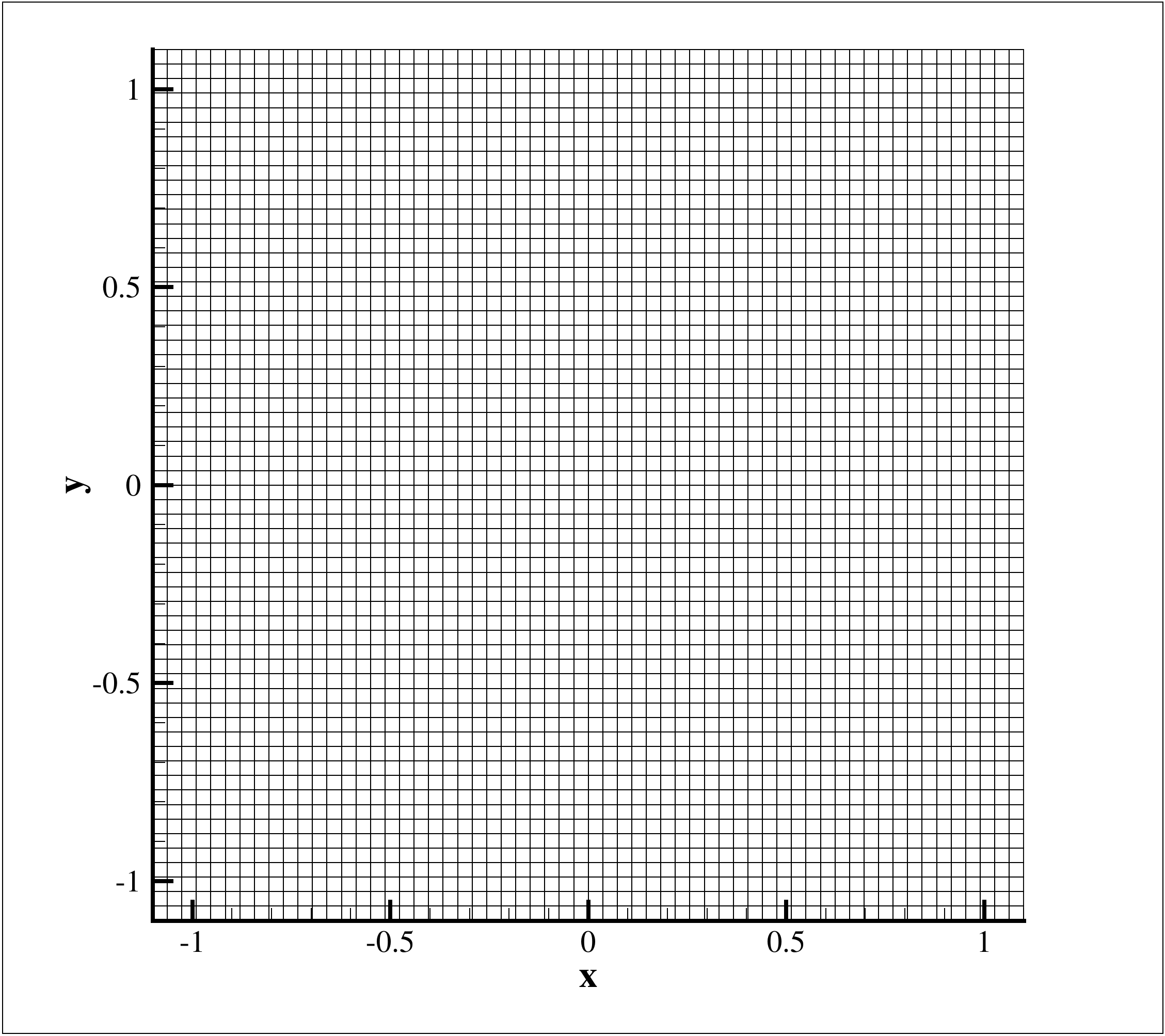} 
  \caption{Initial mesh for the Sedov problem.}
  \label{fig:sedov:mesh:initial}
\end{figure}

The Sedov \cite{sedov1959} problem is a high intensity shock wave propagating outwards, generated by a strong explosion at the origin  due to an energy source.
The material is a gamma-law gas initially at rest with $\gamma = 7/5$.
We consider the computational domain of a square on $[-1.1,1.1]\times[-1.1,1.1]$, with $60\times 60$ cells uniformly distributed, as shown in Fig. \ref{fig:sedov:mesh:initial}.
The initial conditions of the background are given by $(\rho^0,u^0,v^0,p^0) = (1.0, 0,0,1.0\times 10^{-6})$.
An energy spike at the center of the domain is released instantly; this is done by setting the total amount of energy in the cells surrounding the origin.
For the current mesh configuration, we have four quadrilateral elements surrounding the origin point; 
each of them has a volume $\Omega_0$ and each will be assigned an amount of energy $E_{total}$, thus its pressure is 
$p_0 = (\gamma - 1)\rho_0\frac{E_{total}}{\Omega_0}$. 
In this case, $E_{total} = 0.244816$.
With this choice, the shock front of the blast wave will be at radius $r=1$ at time $t = 1$, with a density peak 6.
For this problem, we use the second form of impedance for all three solvers.

\begin{figure}[H]
  \centering
  \subfloat[Burton solver]{
    \includegraphics[trim= 1.0cm 0.7cm 1.0cm 1.5cm,clip,height = 2.5in]
    {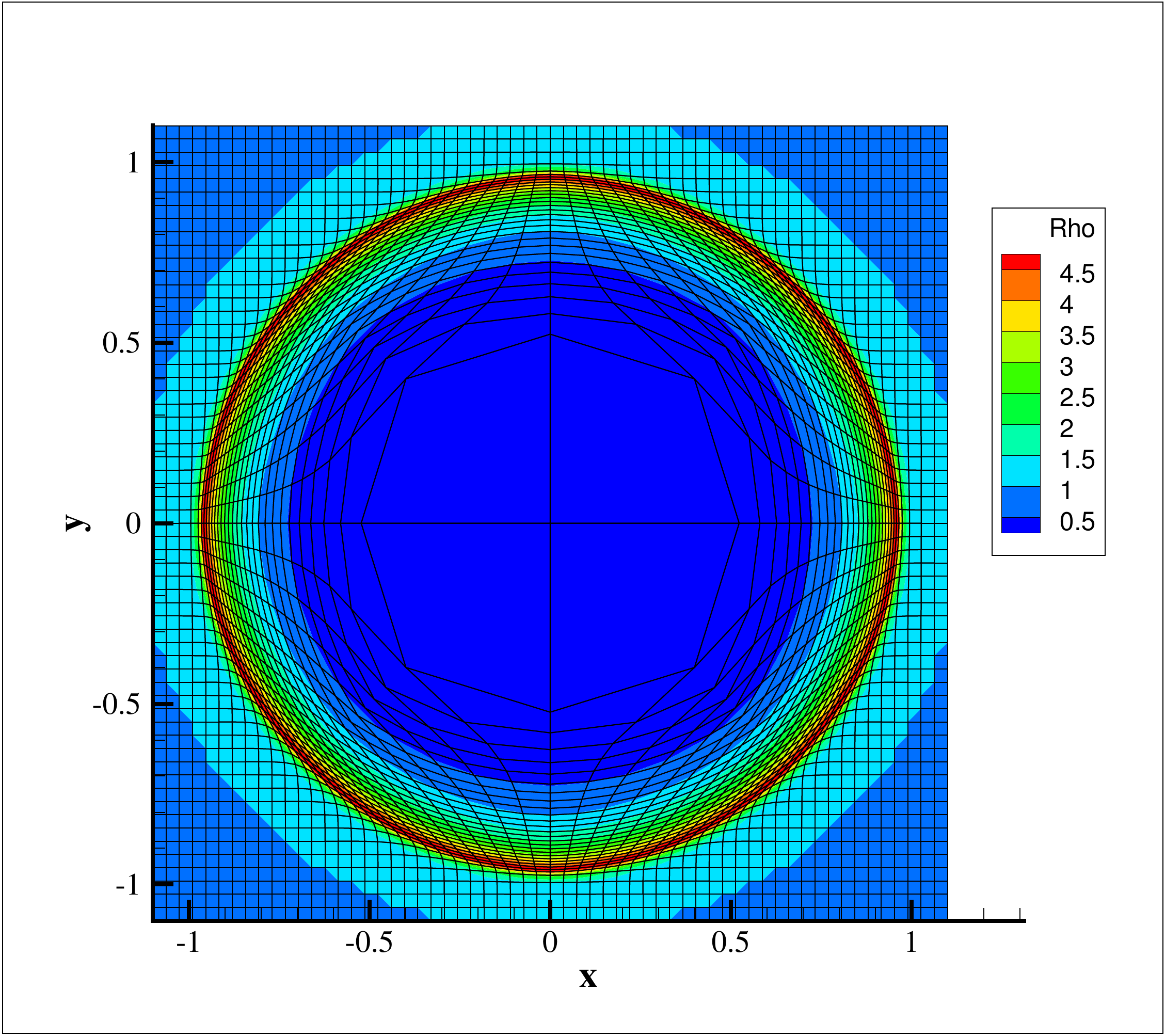} }\\
  \subfloat[Maire solver]{ 
    \includegraphics[trim= 1.0cm 0.7cm 1.0cm 1.5cm,clip,height = 2.5in]
    {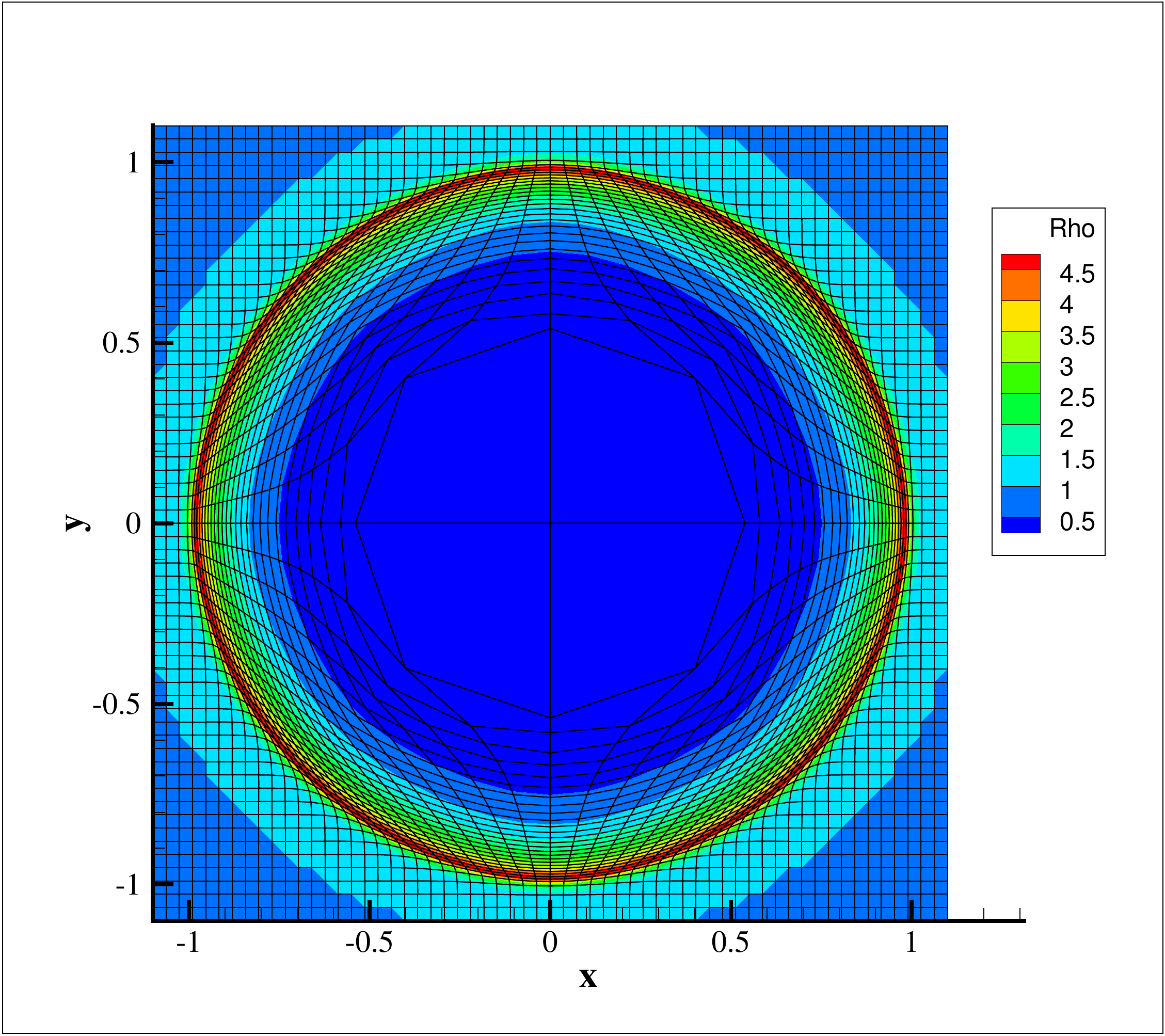} }\\
  \subfloat[LS solver]{ 
    \includegraphics[trim= 1.0cm 0.7cm 1.0cm 1.5cm,clip,height = 2.5in]
    {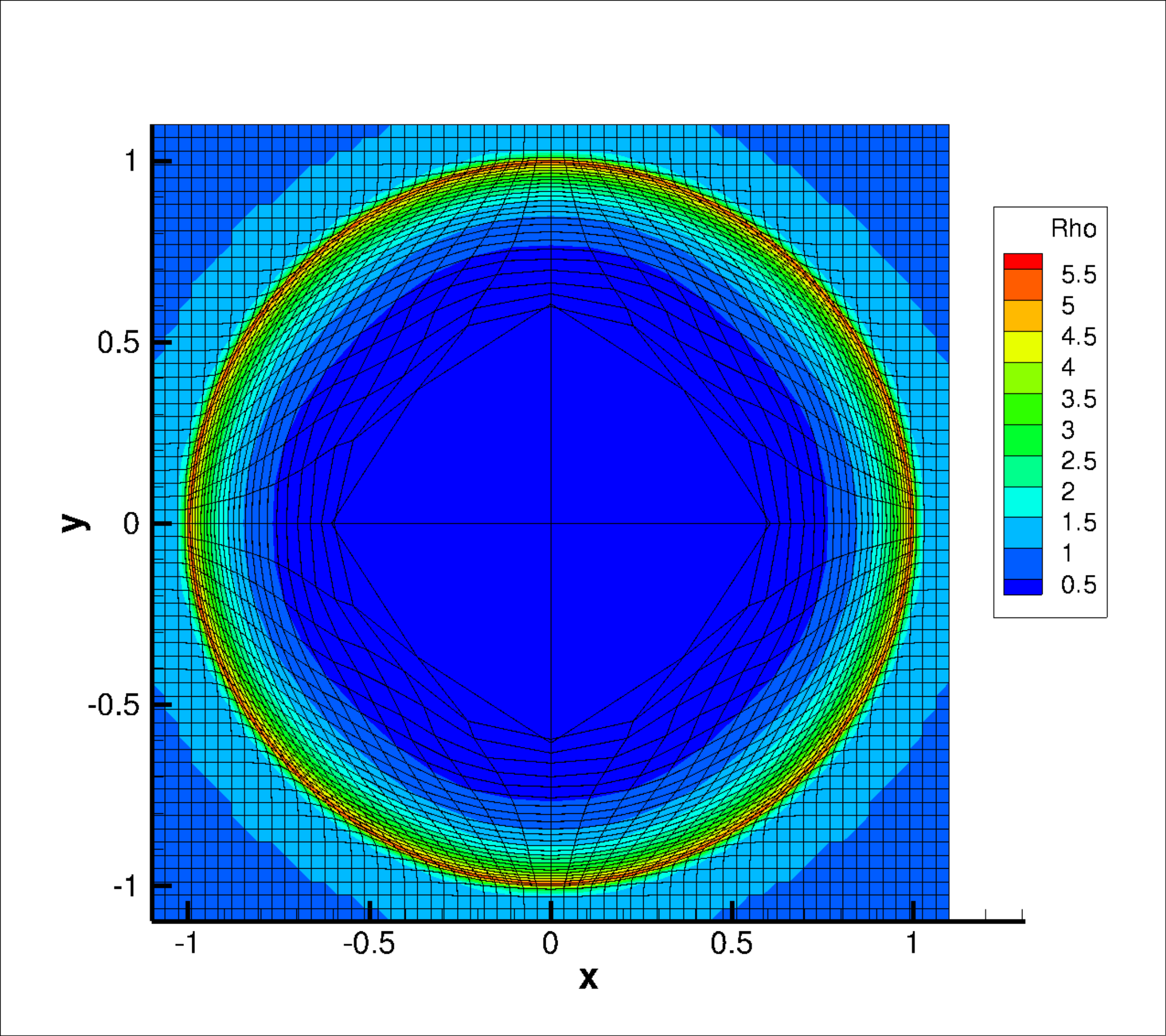} }
  \caption{Final mesh and density contour for the Sedov problem.}
  \label{fig:sedov:densityContour}
\end{figure}

The final mesh and density contours obtained by three solvers are shown in Fig. \ref{fig:sedov:densityContour}.
In the mesh obtained by the LS solver, four nodes on the innermost cells have travelled into other cells.
This is perhaps due to the lack of dissipation of the new solver, compared to the other two.

\begin{figure}[H]
  \centering
  \subfloat[Burton solver]{   
    \includegraphics[trim= 0cm 0cm 0cm 0cm,clip,height = 2.5in]
    {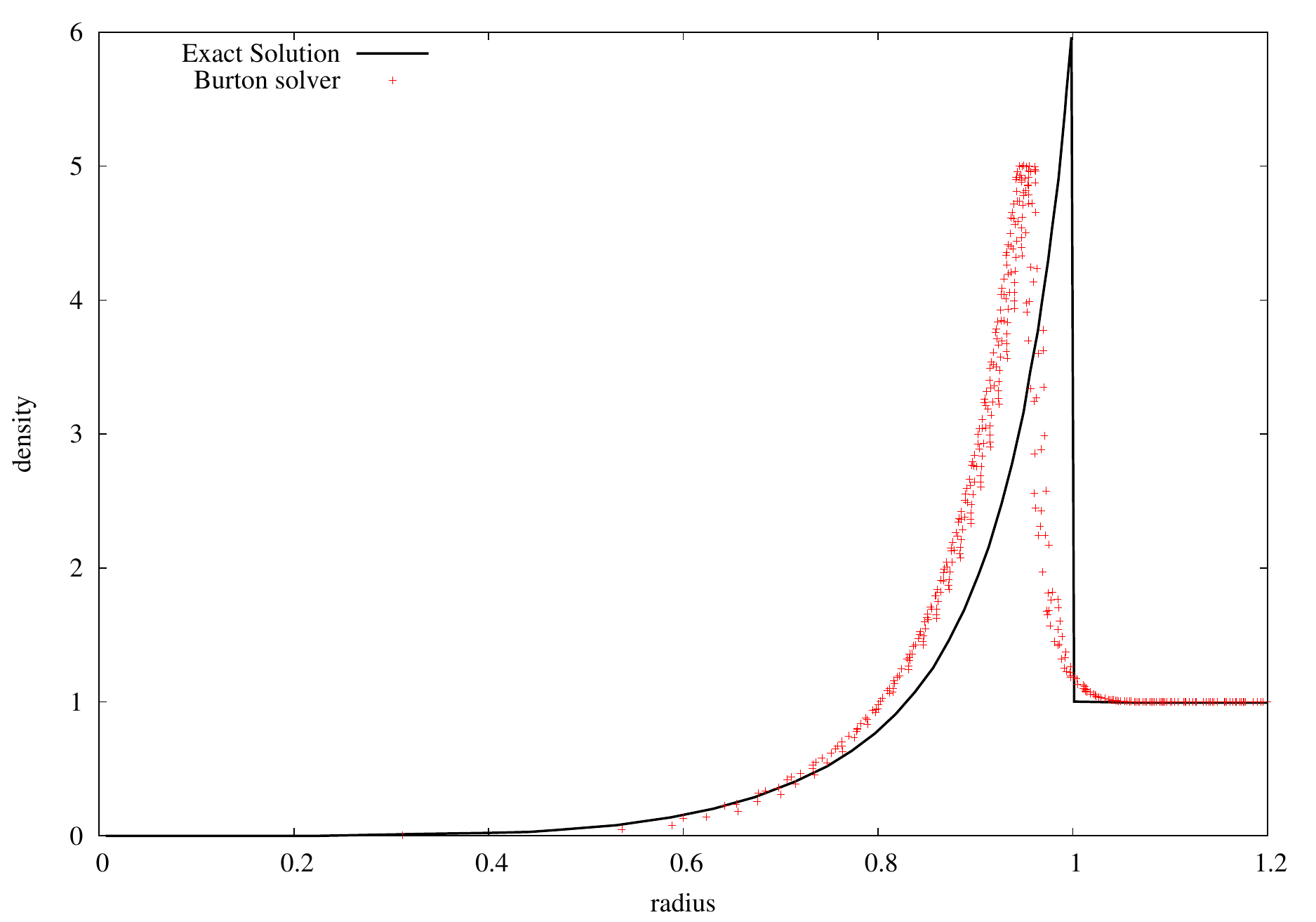}}\\
  \subfloat[Maire solver]{   
    \includegraphics[trim= 0cm 0cm 0cm 0cm,clip,height = 2.5in]
    {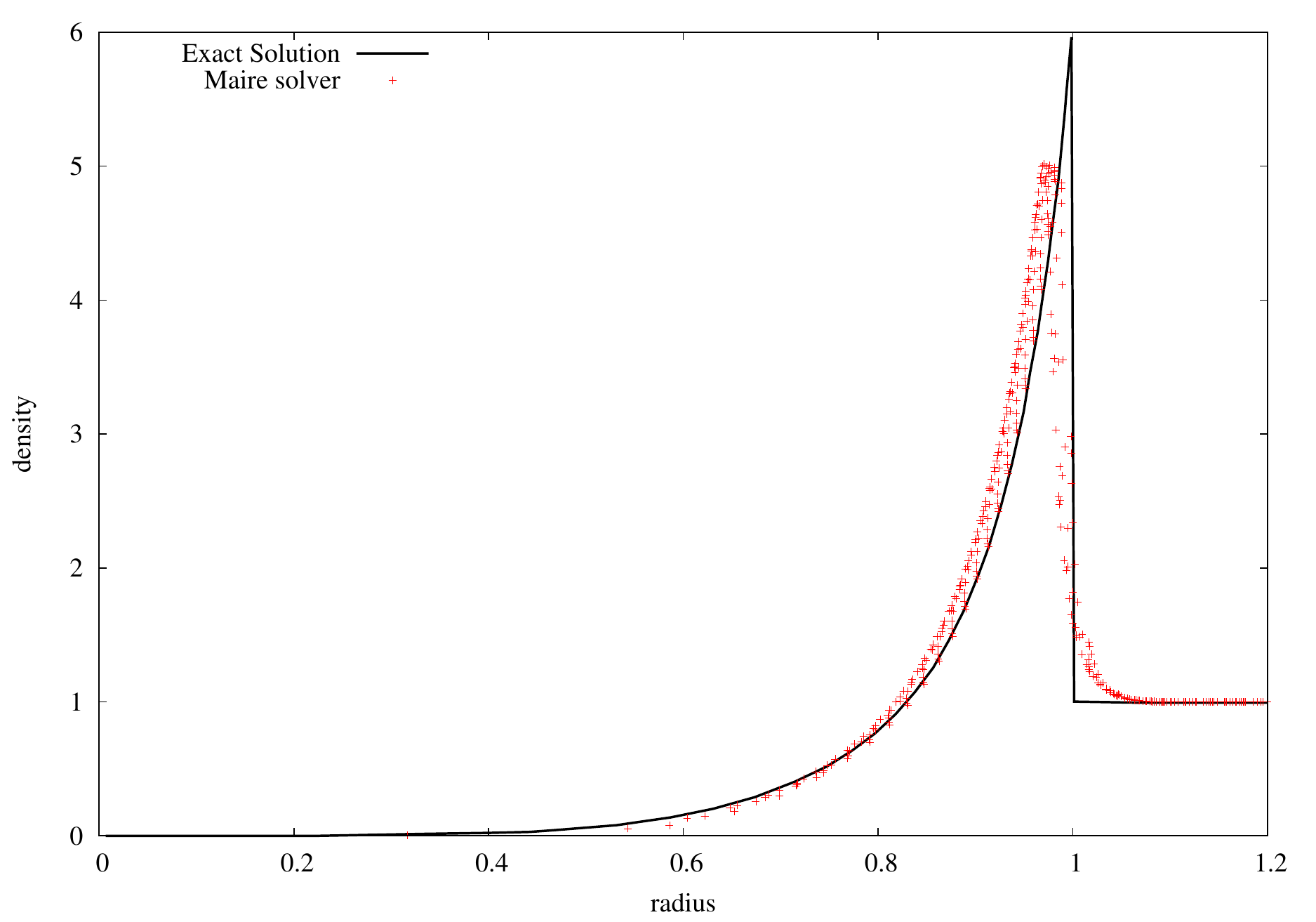}}\\ 
  \subfloat[LS solver]{   
    \includegraphics[trim= 0cm 0cm 0cm 0cm,clip,height = 2.5in]
    {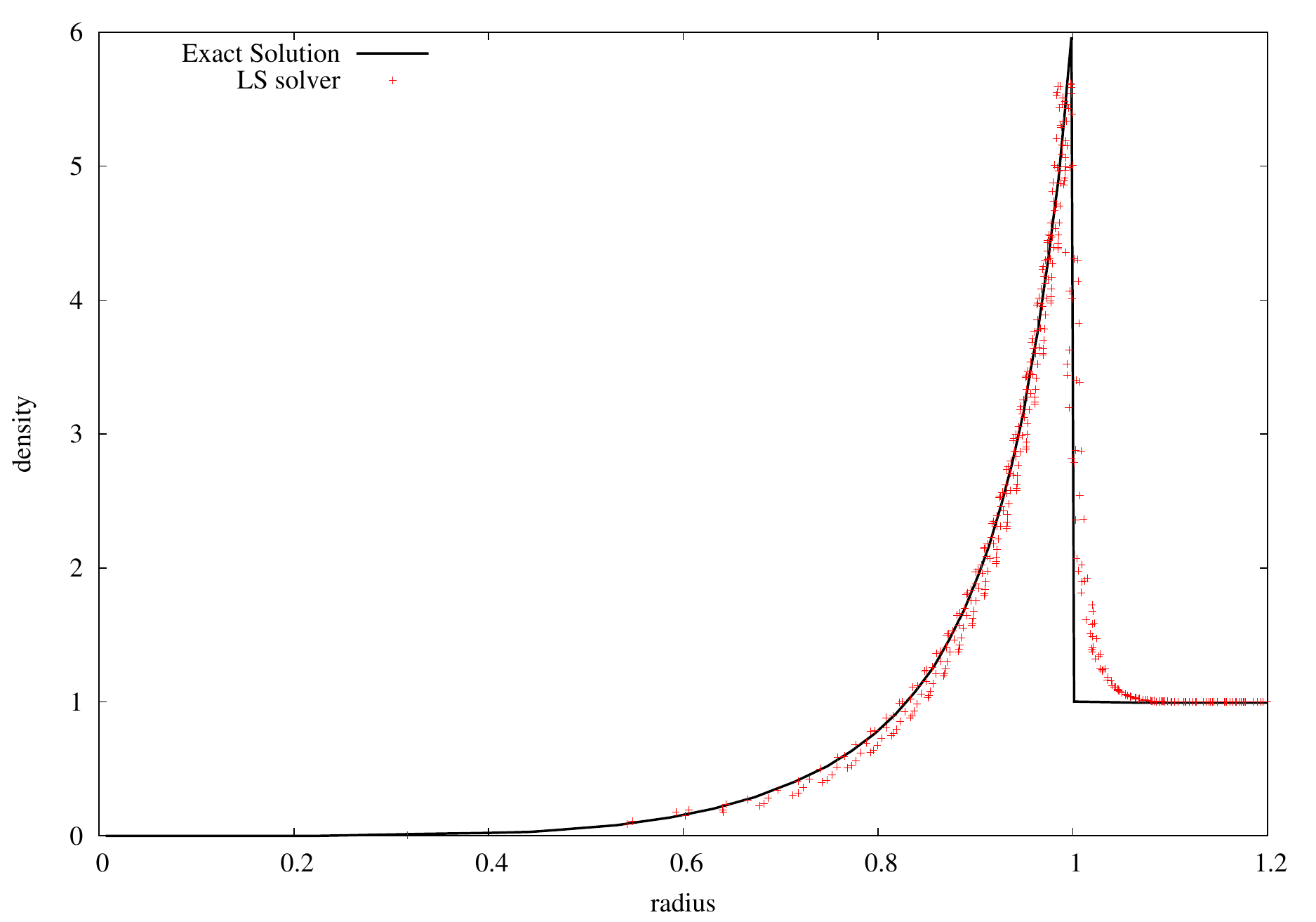}}          
  \caption{Scattered density plots for the Sedov problem.}
  \label{fig:sedov:densityPlot}
\end{figure}

The scattered densities at the cell-centers are plotted against the radius in Fig. \ref{fig:sedov:densityPlot}.
The density profile from the new solver shows a better shock position than that from Burton solver, and has a higher density peak than its counterpart of the Maire solver.

\subsection{Noh Problem}

The Noh problem \cite{noh1987} is a typical test case used extensively in the literature to validate the Lagrangian schemes in the regime of strong shocks.
The material is the ideal gas with $\gamma = 5/3$, with initial density $\rho^0 = 1$ and pressure $p^0 = 1.0\times 10^{-6}$.
The initial velocity is characterized by a unit inward radial vector, i.e., $(u^0,v^0) = (-x/r, -y/r)$ where $r=\sqrt{x^2+y^2}$.
A shock wave generated at the origin due to the converging flow propagates outwards and the density plateau behind it reaches the value 16.

\begin{figure}[H]
  \centering
  \subfloat[global view]{
    \includegraphics[trim= 1.0cm 0.7cm 2.0cm 0.5cm,clip,height = 2.5in]
    {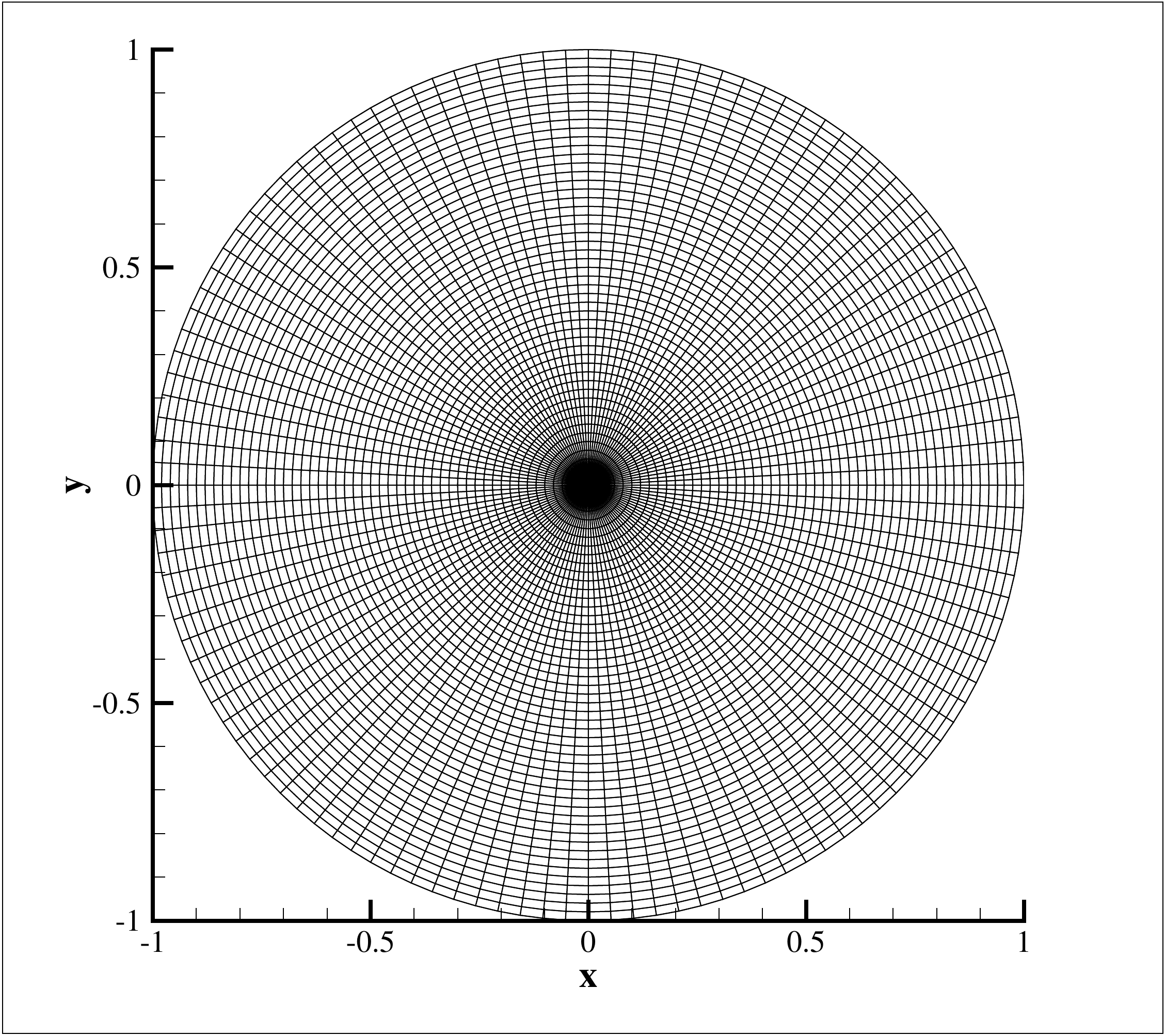} }
  \subfloat[mesh near origin]{
    \includegraphics[trim= 1.0cm 0.7cm 1.0cm 0.5cm,clip,height = 2.5in]
    {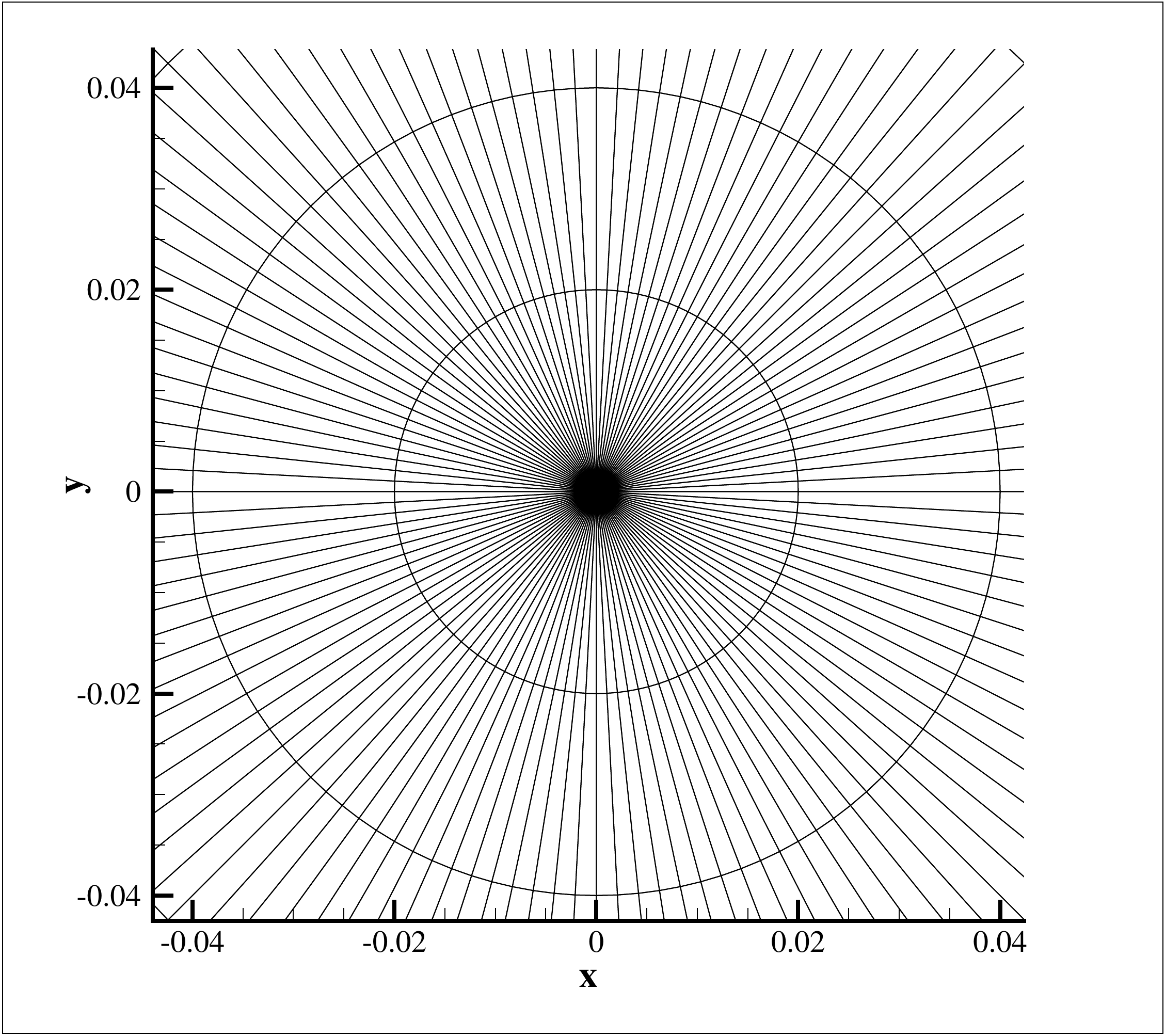} }
  \caption{Initial grids for the Noh problem.}
  \label{fig:noh:mesh:initial}
\end{figure}

For this problem, we use the polar mesh in the computational domain $[0,1]\times[0,2\pi]$, with 50 cells in the radial direction and $30\times 4$ cells in the circumferential direction.
The initial mesh is shown in Fig. \ref{fig:noh:mesh:initial}, where the triangles are surrounding the origin node, and elsewhere are quadrangles.
The simulation time ends at $t=0.6$.

\begin{figure}[H]
  \centering
  \subfloat[density]{
    \includegraphics[trim= 1.0cm 0.7cm 2.5cm 0.5cm,clip,height = 2.5in]
    {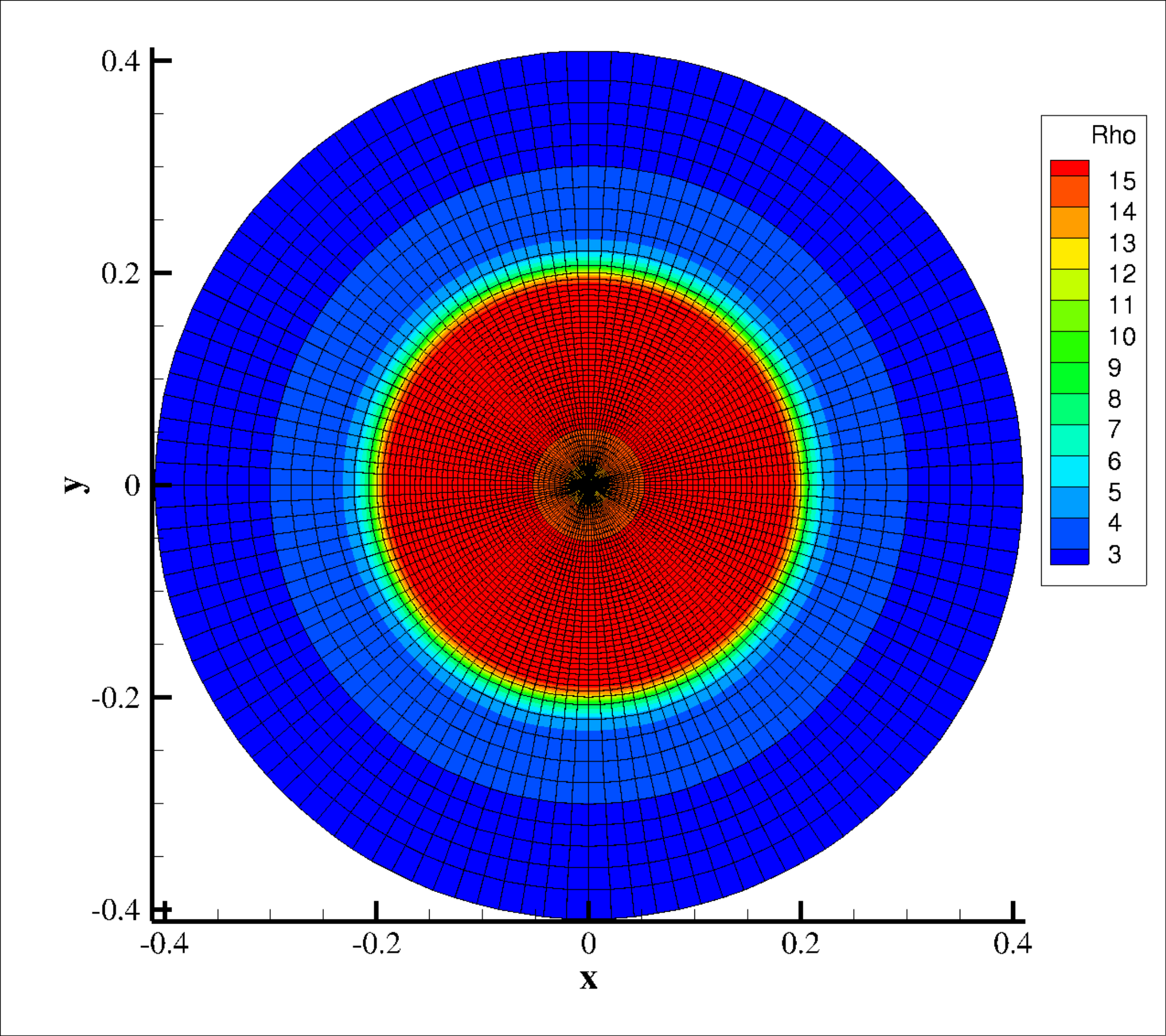} }
  \subfloat[zoomed-in]{ 
    \includegraphics[trim= 1.0cm 0.7cm 0.3cm 0.5cm,clip,height = 2.5in]
    {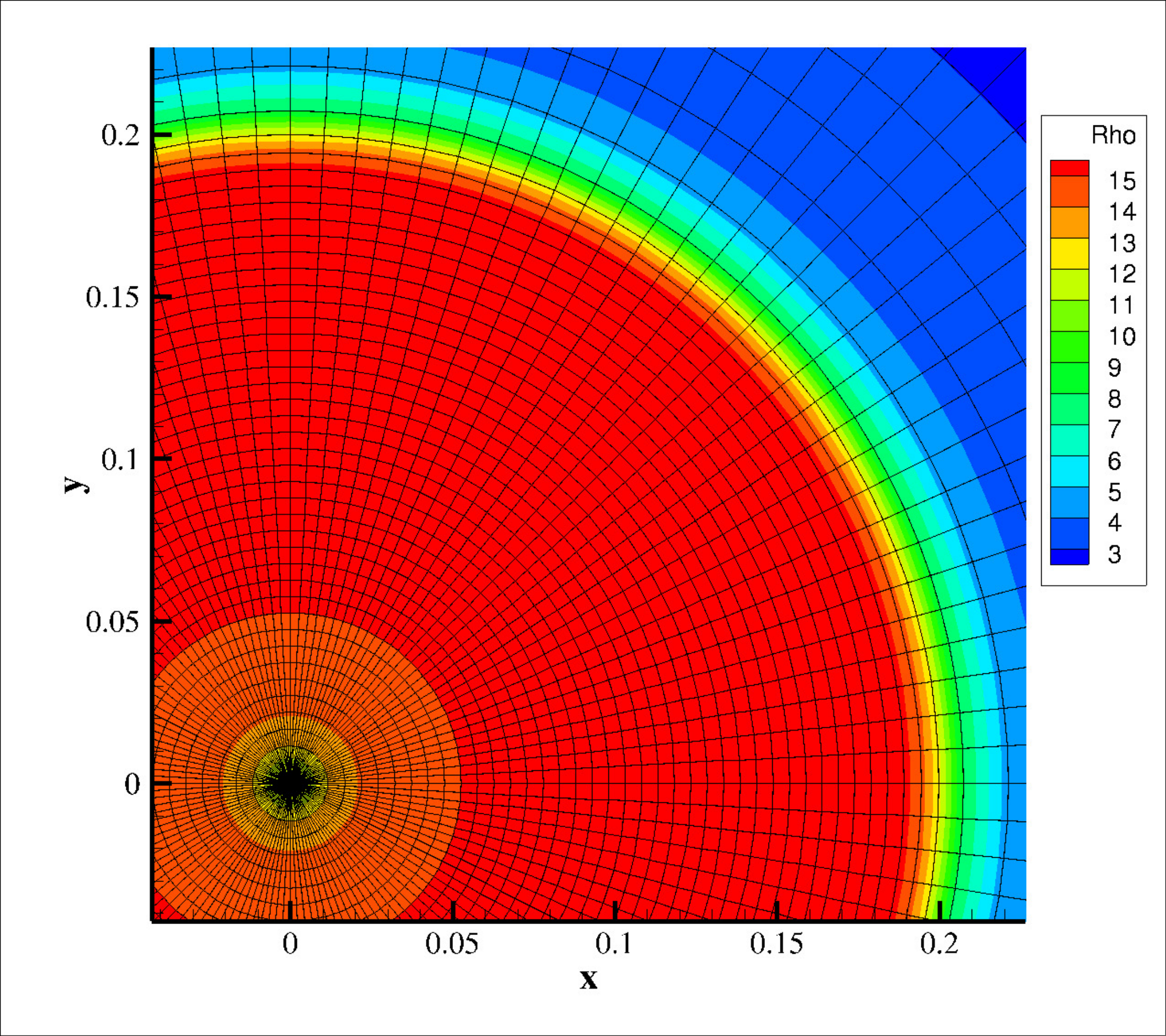} }
  \caption{Final mesh and density contour for the Noh problem (Burton solver).}
  \label{fig:noh:cch:densityContour:p0}
\end{figure}

\begin{figure}[H]
  \centering
  \subfloat[density]{
    \includegraphics[trim= 1.0cm 0.7cm 2.5cm 0.5cm,clip,height = 2.5in]
    {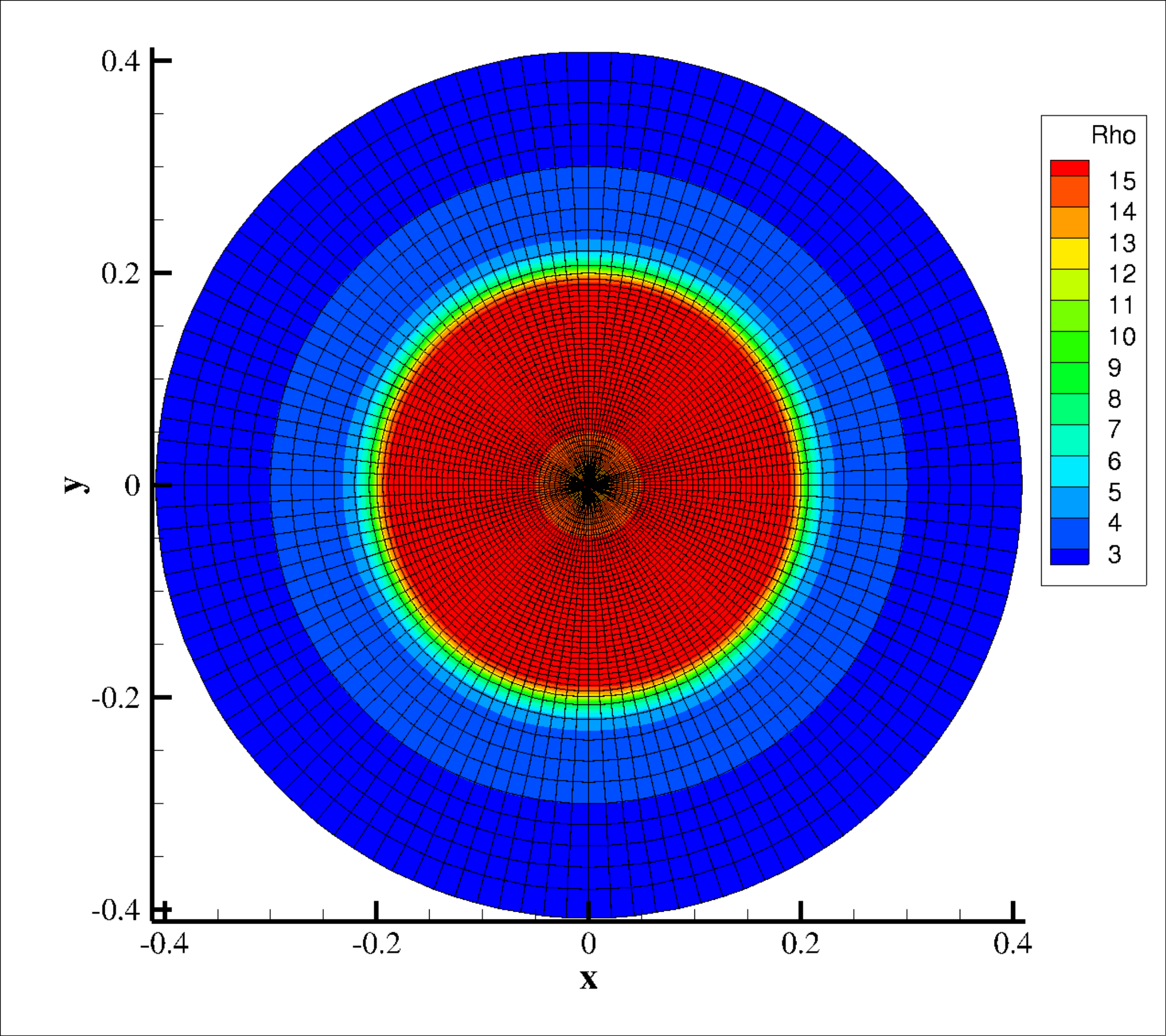} }
  \subfloat[zoomed-in]{ 
    \includegraphics[trim= 1.0cm 0.7cm 0.3cm 0.5cm,clip,height = 2.5in]
    {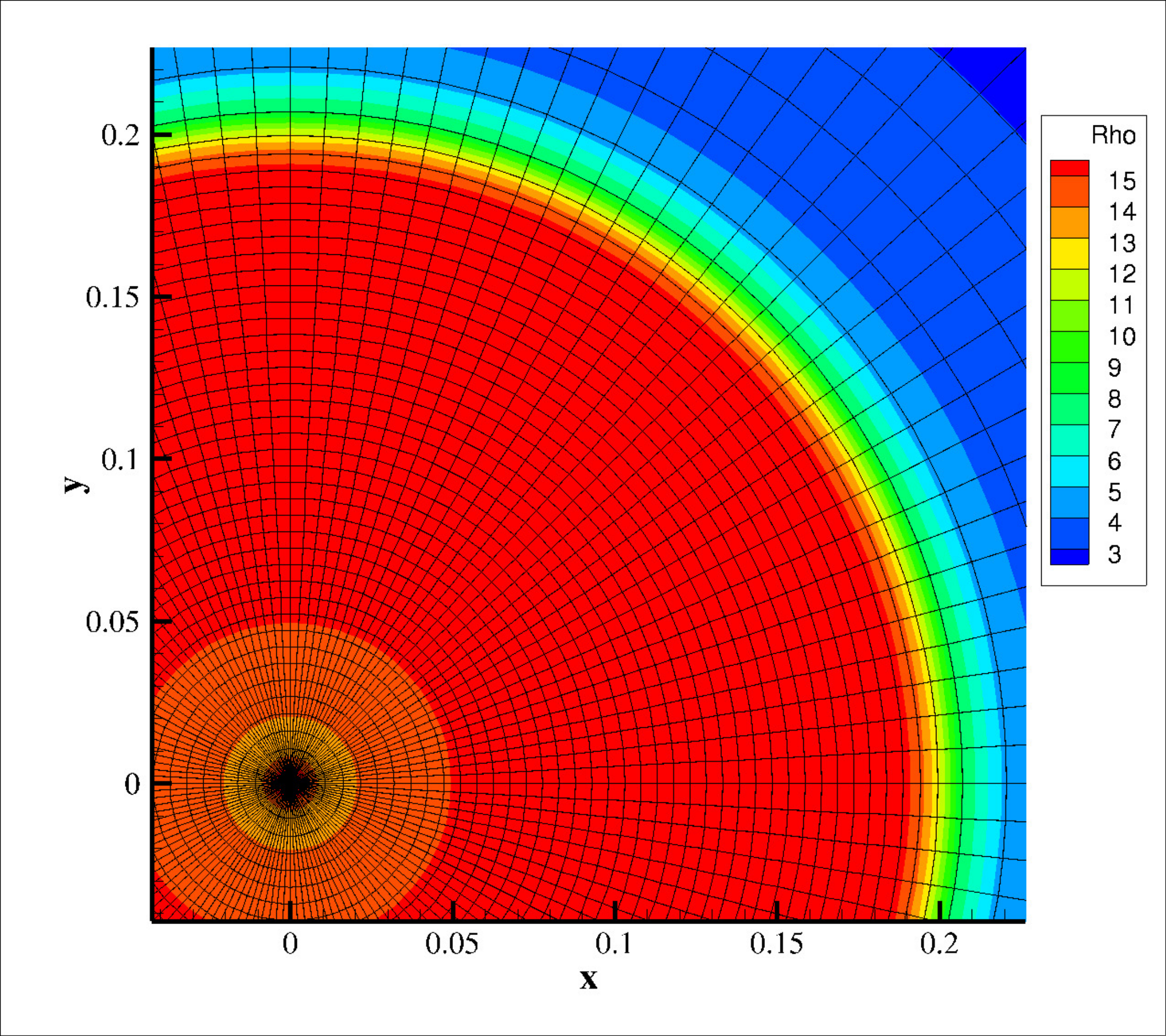} }
  \caption{Final mesh and density contour for the Noh problem (Maire solver).}
  \label{fig:noh:mai:densityContour:p0}
\end{figure}

\begin{figure}[H]
  \centering
  \subfloat[density]{
    \includegraphics[trim= 1.0cm 0.7cm 2.5cm 0.5cm,clip,height = 2.5in]
    {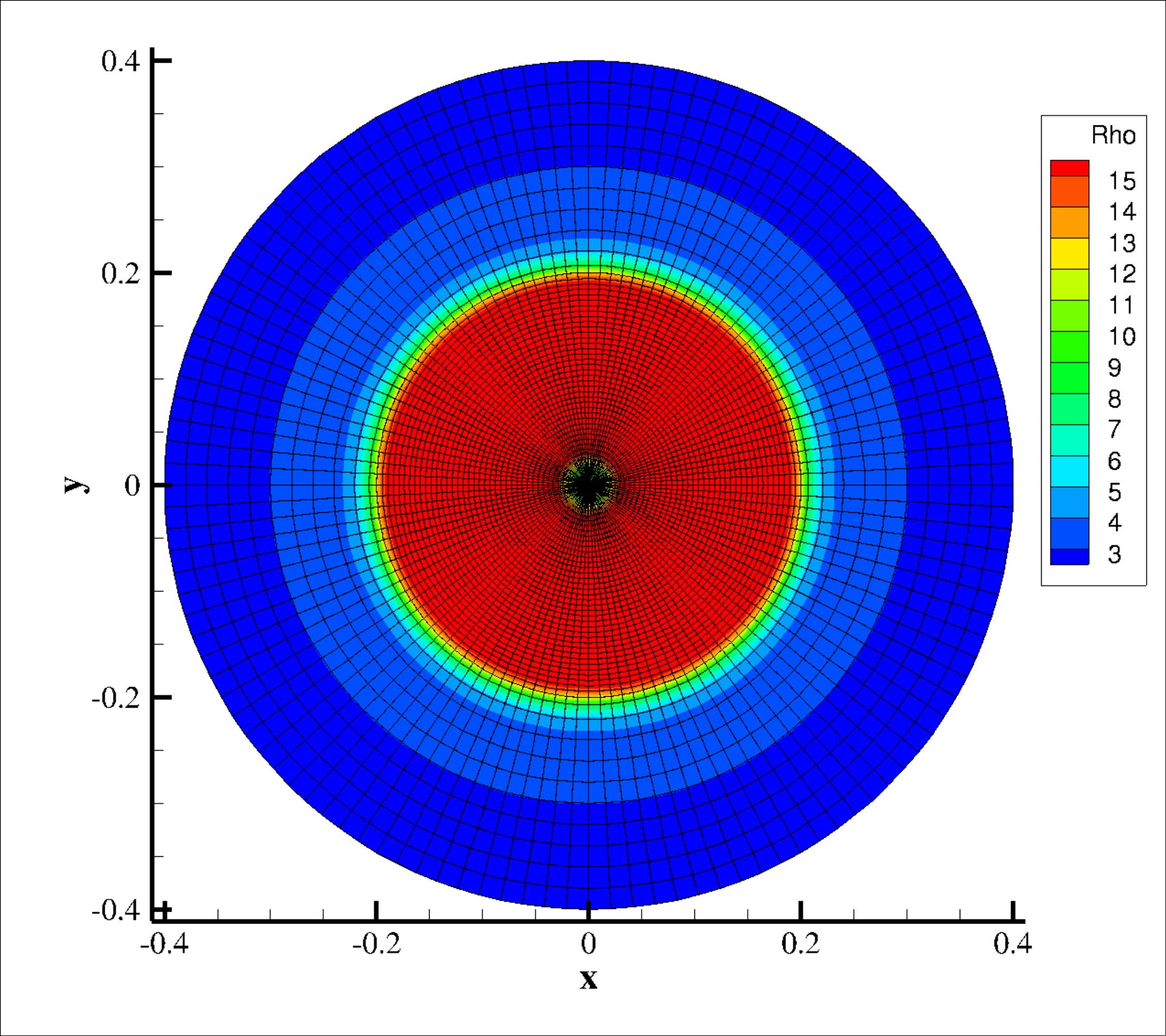} }
  \subfloat[zoomed-in]{ 
    \includegraphics[trim= 1.0cm 0.7cm 0.3cm 0.5cm,clip,height = 2.5in]
    {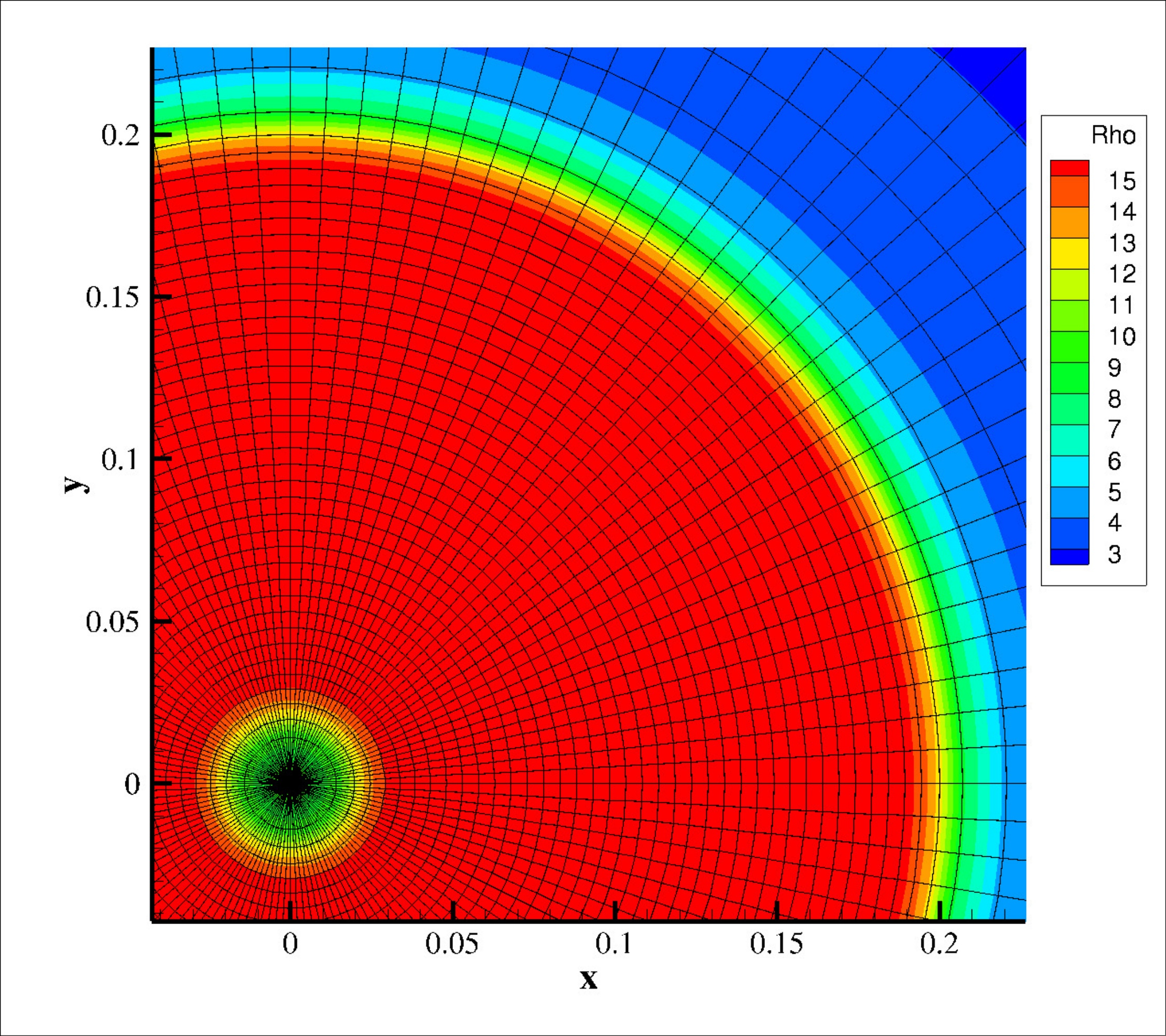} }
  \caption{Final mesh and density contour for the Noh problem (LS solver).}
  \label{fig:noh:ls:densityContour:p0}
\end{figure}

\begin{figure}[H]
  \centering 
  \includegraphics[trim= 0cm 0cm 0cm 0cm,clip,height = 2.5in]
  {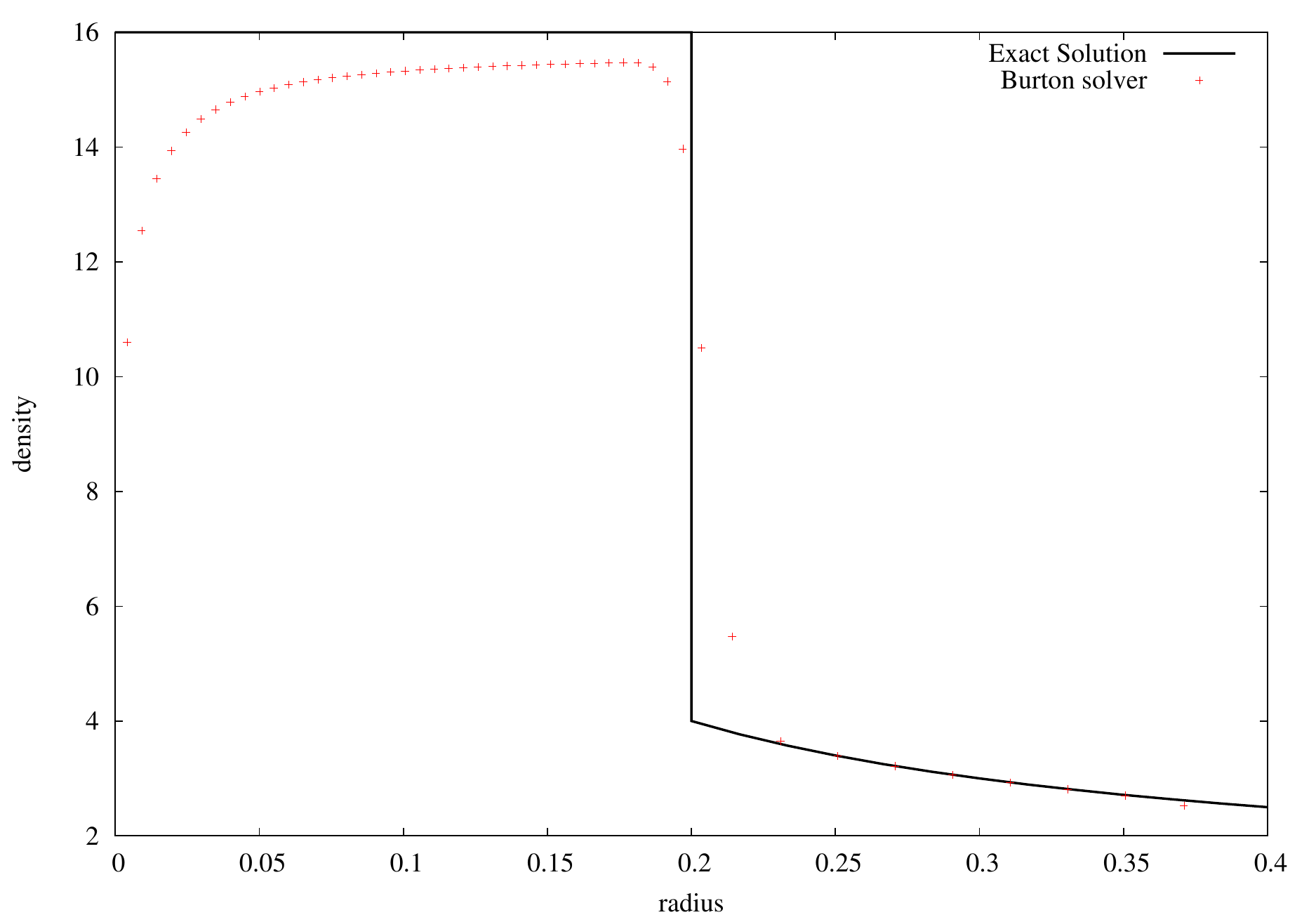} 
  \caption{Scattered density plots for the Noh problem (Burton solver).}
  \label{fig:noh:cch:densityPlot}
\end{figure}

\begin{figure}[H]
  \centering 
  \includegraphics[trim= 0cm 0cm 0cm 0cm,clip,height = 2.5in]
  {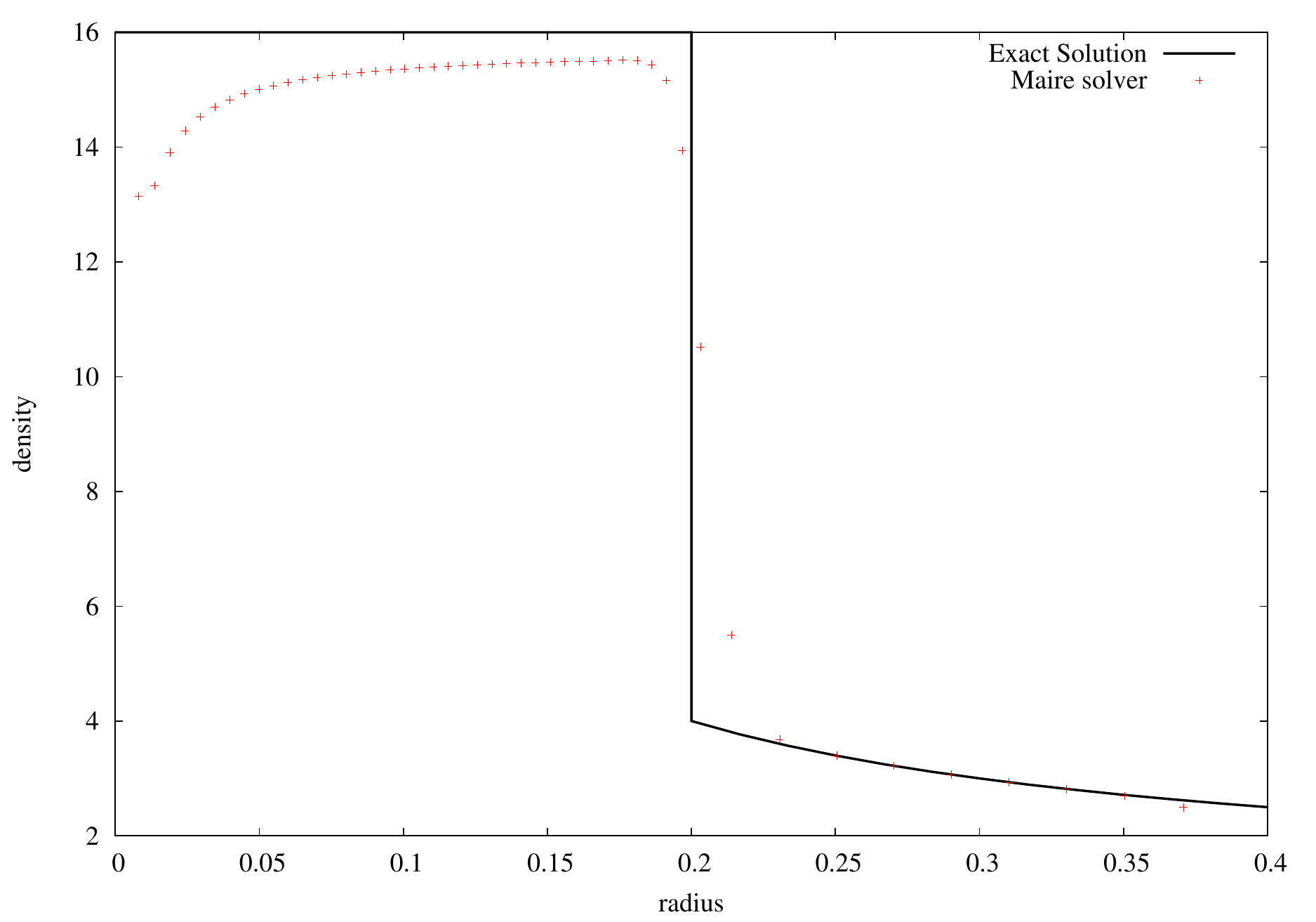} 
  \caption{Scattered density plots for the Noh problem (Maire solver).}
  \label{fig:noh:mai:densityPlot}
\end{figure}

\begin{figure}[H]
  \centering 
  \includegraphics[trim= 0cm 0cm 0cm 0cm,clip,height = 2.5in]
  {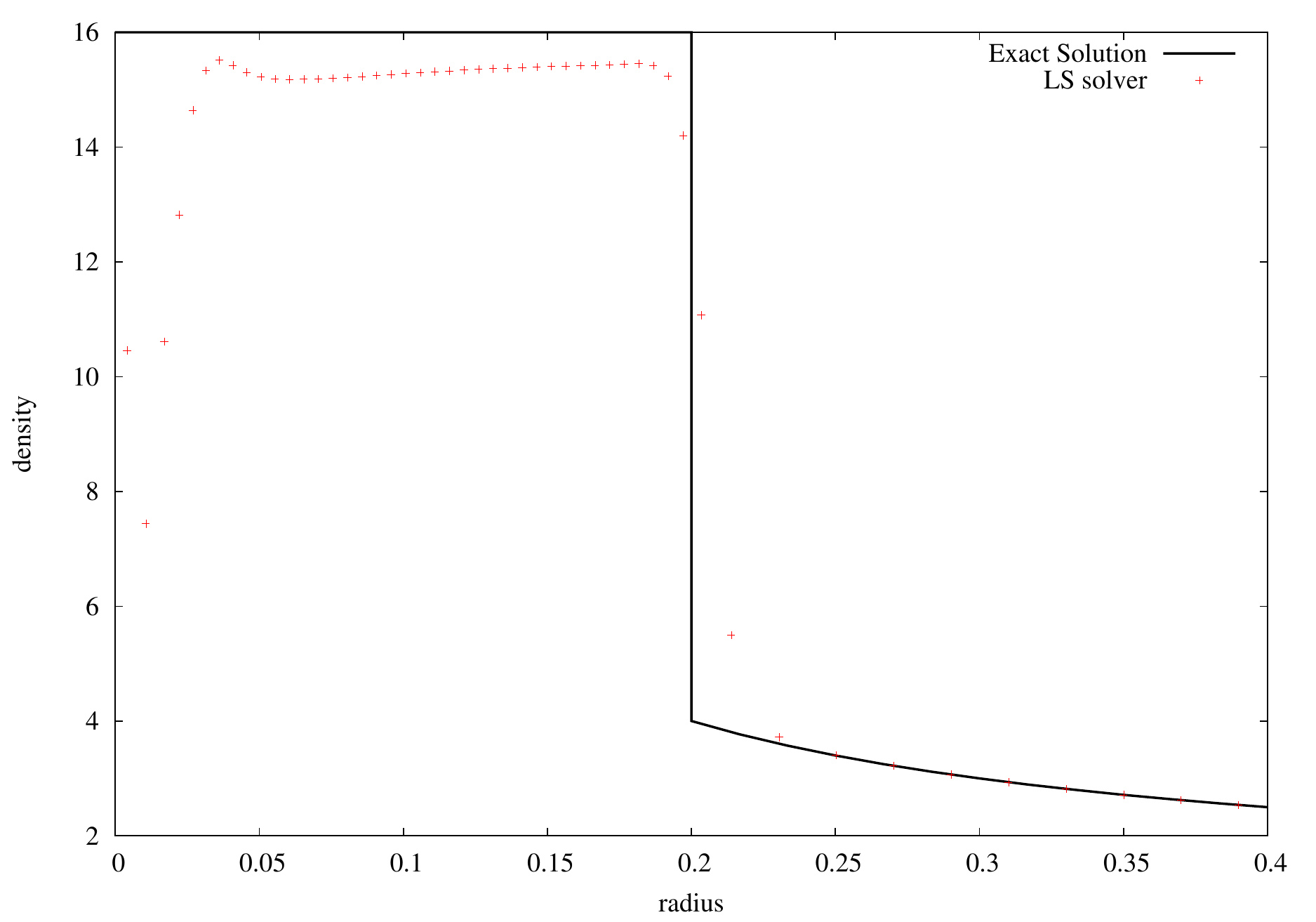} 
  \caption{Scattered density plots for the Noh problem (LS solver).}
  \label{fig:noh:ls:densityPlot}
\end{figure}

The final meshes and density contours are displayed in Fig. \ref{fig:noh:cch:densityContour:p0}, 
Fig. \ref{fig:noh:mai:densityContour:p0} 
and Fig. \ref{fig:noh:ls:densityContour:p0},
and the scattered densities versus radius are plotted in Fig. \ref{fig:noh:cch:densityPlot}, 
Fig. \ref{fig:noh:mai:densityPlot} 
and Fig. \ref{fig:noh:ls:densityPlot}.
We can see there're some difference between the solution from the new LS solver and those from the other two solvers.
This phenomenon is under further investigation.

\section{Conclusions}
\label{sec:conclusion}
A new least-squares based nodal Riemann solver has been proposed to solve the compressible
Euler equations in the updated Lagrangian formulation, where
the conservative variables are solved.
This formulation is the Lagrangian limit of the unsplit ALE formulation, 
by invoking the assumption that the grid velocity is equal to the fluid velocity at cell boundaries.
One feature of the new solver is the single Riemann pressure at a node, which together with the Riemann velocity, are obtained by solving a pressure-velocity coupled least-squares system.
The resulting nodal solutions are used to move the mesh as well as evaluate the numerical flux at cell interface.  
A number of benchmark test cases have been set up to assess its accuracy and stability.
The performance of the new solver are compared with that from
two other acoustic solvers developed by Burton et al. and Maire et al., respectively. 
The extension of this method to higher order is quite straightforward and is under investigation.

\section*{References}

\bibliography{mybibfile}

\end{document}